\documentclass[useAMS,usenatbib]{mn2e}

\usepackage{amssymb}
\usepackage{amsfonts}
\usepackage{mncite}
\usepackage{epsfig}
\usepackage{psfig}

\begin{document}

\title[Disc self-gravity and star formation]
{The role of disc self-gravity in the formation of protostars and protostellar discs}

\author[W.K.M. Rice, J.H. Mayo \& Philip J. Armitage]
 {W.K.M. Rice$^1$\thanks{E-mail: wkmr@roe.ac.uk}, J.H. Mayo$^1$, Philip
 J. Armitage$^{2,3}$ \\
$^1$ SUPA\thanks{Scottish Universities Physics Alliance},
Institute for Astronomy, University of Edinburgh, Blackford Hill, Edinburgh, EH9 3HJ \\
$^2$ JILA, Campus Box 440, University of Colorado, Boulder CO 80309 \\
$^3$ Department of Astrophysical and Planetary Sciences, University of
 Colorado, Boulder CO 80309 \\}

\maketitle

\begin{abstract}
We use time-dependent, one-dimensional disc models to investigate 
the evolution of protostellar discs that form through the collapse of
molecular cloud cores and in which the primary transport mechanism is self-gravity.
We assume that these discs settle into a state of thermal equilibrium with $Q = 2$
and that the strength of the angular momentum transport is set by the cooling rate of the disc. The 
results
suggest that these discs will attain a quasi-steady state that persists for a number
of free-fall times and in which most of the mass within $100$ au is located inside $10-20$ au.
This pile-up of mass in the inner disc could result in temperatures that are high enough
for the growth of MHD turbulence which could rapidly drain the inner disc and 
lead to FU Orionis-like outbursts. In all our simulations, the inner regions of
the discs ($r < 40$ au) were stable against fragmentation, while fragmentation was possible
in the outer regions ($r > 40$ au) of discs that formed from cores that had enough initial angular 
momentum to deposit sufficient mass in these outer regions. The large amounts of mass in these
outer regions, however, suggests that fragmentation will lead to the formation of sub-stellar
and stellar mass companions, rather than planetary mass objects.  Although mass accretion rates
were largely consistent with observations, the large disc masses suggest that an additional
transport mechanism (such as MRI occuring in the upper layers of the disc) must operate
in order to drain the remaining disc material within observed disc lifetimes.
\end{abstract}


\section{Introduction}
The formation of low-mass stars occurs through the collapse of cold, dense molecular cloud
cores \citep{terebey84}.  These cores, however, generally contain amounts of angular momentum
far in excess of the rotational angular momentum of a single star \citep{caselli02}. Most
of the mass must therefore first pass through a circumstellar disc before accreting onto
the central star. One of the open problems in star and planet formation is what mechanism
acts to transport the angular momentum outwards, allowing this accretion to take place. 

It is now generally accepted that in most astrophysical discs, angular momentum transport is
driven by magnetohydrodynamic (MHD) turbulence initiated by the magneto-rotational instability
(MRI) \citep{balbus91,papaloizou03}. During the earliest stages of star formation protoplanetary
discs are, however, cold and dense and probably do not have even the very small degree
of ionization needed to sustain MHD turbulence \citep{blaes94}.  At these early times, the
disc is, however, likely to be massive and disc self-gravity may then provide an alternate and
possibly dominant transport mechanism through the growth of the gravitational instability
\citep{toomre64,lin87,laughlin94}.

Although it has been suggested that gravitationally unstable discs may fragment to form gas
giant planets or substellar companions \citep{boss98,boss02}, it is now generally accepted that the 
conditions for fragmentation are difficult to achieve in the inner planet forming
regions of protostellar discs \citep{matzner05,rafikov05,boley06,whitworth06,stamatellos08,
forgan09}. A self-gravitating protostellar disc is more likely to settle into a quasi-steady
state in which the instability acts to transport angular momentum outwards \citep{gammie01,
rice03, lodato04, vorobyov07}. It has been shown \citep{lodato04,lodato05} 
that in this quasi-steady state,
the transport can be approximated as a local viscous process for all but the most
massive discs.  Since the disc will also be in thermal equilibrium, the strength of the
angular momentum transport is also set by the cooling rate of the disc \citep{gammie01,rice03}.
Using the above results, \citet{rice09} have shown that if self-gravity is the dominant
transport mechanism, protostellar discs will settle into quasi-steady states that are
largely independent of the initial conditions. The resulting surface density
profiles are quite steep with $\sim 80$ \% of the mass within $50$ au located
inside $10 -20$ au. \citet{rice09} also found that the quasi-steady mass accretion
rates depend strongly on the disc mass and on the mass of the central star. Their simulations
also suggested that no regions of the disc (which extended to $50$ au) were susceptible to
fragmentation with the inner $10 - 20$ au being particularly stable, consistent
with other recent calculations \citep{clarke09,rafikov09}.

The simulations carried out by \citet{rice09} assumed initial power-law surface density profiles
that were evolved until a quasi-steady state - which we define as a state in which
the mass accretion rate is approximately independent of radius - was achieved. Although this gives information
about the quasi-steady nature of such systems, it doesn't tell us if such a state can
be achieved under realistic conditions. Here, we extend the work of \citet{rice09} by 
considering the evolution of self-gravitating, circumstellar discs 
formed through infall from a molecular cloud core.  We find that these disc do settle
rapidly into quasi-steady states, with properties similar to that found by \citet{rice09}. 
The mass accretion rates are also generally consistent with observations \citep{muzerolle05},
although at early times the simulated accretion rates are somewhat higher than observed.  Although the disc
masses are generally higher than observed, a significant fraction of the mass is located
in the optically thick inner regions and would not be easily detected using current techniques. 
These results are consistent with the suggestion \citep{armitage01,zhu09} that
mass will pile-up in the inner regions of the disc and will drain episodically 
 onto the
star - potentially explaining FU Orionis outbursts \citep{hartmann96} - 
when the temperature becomes sufficiently high for MRI to operate.

The paper is organised as follows. In Section 2, we describe the
basic model that builds the disc from an infalling molecular
cloud core and evolves the disc self-consistently through,
primarily, self-gravity.  In Section 3, we describe the results,
and in Section 4 we summarize our conclusions.

\section{Basic Model}

\subsection{Infall}
We base the model here on that of \citet{lin90} which itself is based on the earlier work of \citet{cassen81}. 
We consider a molecular cloud core of mass $M_c$, radius $R_c$ and angular velocity $\Omega_c$ that is assumed to have an
initially uniform density, $\rho_c$.  The core is allowed to collapse under gravity, with the collapse occuring on 
a characteristic free-fall time of $t_{\rm ff} = (3 \pi/32 G \rho_c)^{1/2}$.  As in \citet{lin90}, we assume that
the collapse is initially steady \citep{larson69} and then, after one free-fall time, decreases smoothly to zero over 
a further free-fall time. The form of the accretion rate is therefore \citep{lin90}
\begin{equation}
\dot{M}(t) = 
\left\{
\begin{array}{ll}
\frac{M_c}{t_{\rm ff}}  &0  \le  t \le t_{\rm ff}\\
\frac{1}{2}\frac{M_c}{t_{\rm ff}} \left[1 + \cos \frac{\pi (t - t_{\rm ff})}{t_{\rm ff}}\right]  &t_{\rm ff} \le t \le 2 t_{\rm ff} \\
0  &2t_{\rm ff} \le  t  
\end{array}
\right\},
\label{accr_rate}
\end{equation}
which implies that the total amount of mass accreted is actually $1.5 M_c$.
The infalling material is assumed to conserve angular momentum and consequently produces a centrifugally supported disc which, in 
the absence of angular momentum transport, would have a maximum radius of $a_{\rm max} = \Omega_c^2 R_c^4/G M_c$.  We model
the infall \citep{cassen81} as the collapse of shells with each part of a shell striking the disc plane at the same time.
The fraction of he mass of a shell, with initial radius $r_o$ and enclosed mass $M(<r_o)$, that strikes the disc within radius $r$ is \citep{lin90}
\begin{equation}
g(r) = 1 - \left(1 - \frac{r}{a_o}\right)^{1/2}   0 \le r \le a_o,
\label{shell_frac}
\end{equation}
where
\begin{equation}
a_o = \frac{\Omega_c^2 r_o^4}{G M(<r_o)}.
\label{out_rad}
\end{equation}

At $t = 0$ the initial timestep, $dt$, determines, through equation (\ref{accr_rate}), the amount of mass
added and hence the amount of mass in the current shell.  Since the core is assumed to have a uniform density, 
$\rho_c$, this determines the radial extent of the initial shell. Equation (\ref{shell_frac}), together with equation (\ref{out_rad}),
then determines how the mass in the shell is distributed in the disc. In subsequent timesteps, the radial extent of the next shell is 
then determined by ensuring that the amount of mass in the shell matches the amount of mass being added in that timestep, and the mass in 
that shell is distributed in the same way as in the first timestep. 

\subsection{Disc evolution}
In order for the disc to evolve there needs to be some mechanism for transporting angular momentum outwards, allowing mass to accrete
onto the central star.  Generally this is assumed to take the form of some kind of kinematic viscosity, $\nu$, which would cause
an axisymmetric disc of surface density $\Sigma(R,t)$ to evolve as \citep{lynden74,pringle81}
\begin{equation}
\frac{\partial \Sigma}{\partial t} = \frac{3}{r} \frac{\partial}{\partial r} 
\left[ r^{1/2} \frac{\partial}{\partial r} \left(\nu \Sigma r^{1/2} \right) \right].
\label{visc_evolution}
\end{equation}

In the work of \citet{lin90} the kinematic viscosity is assumed to be the sum of two components. The first component is a standard ``turbulent" viscosity
given by
\begin{equation}
\nu_{\rm ss} = \alpha_{\rm ss} \frac{c_s^2}{\Omega},
\label{nu_turb}
\end{equation}
where $c_s$ is the disc sound speed, $\Omega$ is the angular frequency, and $\alpha_{\rm ss} << 1$ is a parameter that determines the efficiency
of the angular momentum transport \citep{shakura73}.  In astrophysical discs, the mechanism that provides this kinematic viscosity is generally
accepted to be magnetohydrodynamic (MHD) turbulence initiated by the magneto-rotational instability \citep{balbus91}. 
The second component, $\nu_{\rm g}$, was only non-zero if the disc was self-gravitating.  

An accretion
disc can become gravitational unstable if \citep{toomre64} 
\begin{equation}
Q = \frac{c_s \kappa}{\pi G \Sigma} \sim 1,
\label{Q}
\end{equation}
where $\kappa$ is the epicyclic frequency and can be replaced by the angular frequency, $\Omega$, in a Keplerian disc.   
Following \citet{lin87}, \citet{lin90} assumed that $\nu_{\rm g}$ was zero unless $Q \le Q_c$, where $Q_c$ was a
critical value below which the disc became gravitationally unstable, and that $\nu_{\rm g}$ increased as $Q$ decreased.  The evolution
of self-gravitating discs has, however, been studied in quite some detail recently \citep{gammie01, rice03, lodato04, mejia05, durisen07}
and it is now generally thought that such discs will tend to settle into a state with constant $Q$ and in which
the heating and cooling rates balance. 

This work therefore differs from \citet{lin90} in two ways.  We firstly assume - at least during
the earliest phases of star formation - that in general the disc is too cold and dense to have even the small degree
of ionization needed to sustain MHD turbulence \citep{blaes94}. In this case $\alpha_{\rm ss} = 0$ and disc self-gravity
is the dominant mechanism for transporting angular momentum. We secondly assume that the disc will quickly settle into quasi-steady state
with $Q \sim 2$.  In this state the disc is assumed to be in thermal equilibrium with the radiative cooling balanced by heating 
occuring primarily through compressions and shocks. If we assume that the energy is dissipated locally, which appears to be
the case when $Q \sim 1$ \citep{balbus99,lodato04}, the dissipation is equivalent to that occurring through an effective gravitational viscosity. 
Matching the dissipation rate with the known cooling rate allows the effective gravitational $\alpha$ to be determined \citep{rice09}
and the disc is then evolved using equation (\ref{visc_evolution}).

\subsection{Determining the effective gravitational viscosity}
If we assume that a quasi-steady, self-gravitating disc introduces an effective gravitational viscosity, $\nu_g$, the
dissipation in the disc, $D(R)$, per unit area per unit time is \citep{bell94}
\begin{equation}
D(R) = \frac{9}{4} \nu_g \Sigma \Omega^2.
\label{diss}
\end{equation}
To determine the cooling rate we need to know the midplane temperature and the optical depth. The midplane temperature
can be determined using
\begin{equation}
T_c = \frac{\mu m_{\rm p} c_s^2}{\gamma k_{\rm B}},
\label{cent_T}
\end{equation}
where $\mu = 2.4$ is the mean molecular weight, $m_{\rm p}$ is the proton mass, $\gamma = 1.4$ is the specific heat ratio,
$k_{\rm B}$ is Boltzmann's constant, and $c_s$ is the sound speed determined from the condition that $Q = 2$. We have the 
additional constraint that $T_c$ cannot be less than $10$ K.  We approximate the optical depth using
\begin{equation}
\tau = \int_0^\infty dz \kappa ( \rho_z,T_z ) \rho_z \approx H \kappa ( {\bar{\rho}}, \bar{T} )
\bar{\rho},
\label{tau}
\end{equation}
where $\kappa$ is the Rosseland mean opacity and ${\bar{\rho}}$ and ${\bar{T}}$ are an average density
and temperature for which we use ${\bar{\rho}} = \Sigma/(2 H)$ and ${\bar{T}} = T_c$.  The Rosseland
mean opacities are determined using the analytic approximations from \citet{bell94}. Once the optical
depth is known, the cooling function, $\Lambda$, can be approximated using \citep{hubeny90} 
\begin{equation}
\Lambda = \frac{16 \sigma}{3} (T_c^4 - T_o^4) \frac{\tau}{1 + \tau^2},
\label{coolfunc}
\end{equation}
where $T_o = 10$ K is an assumed minimum temperature set by some background sources \citep{stamatellos07},
and the last term is introduced to smoothly interpolate between optically thick and optically thin regions
\citep{johnson03}.  The effective gravitational viscosity can then be determined by equating equations
(\ref{diss}) and (\ref{coolfunc}). If, however, we rewrite the gravitational viscosity as  $\nu_g = \alpha_g c_s^2 / \Omega$ and
use that $t_{\rm cool} = U/\Lambda$, where $U$ is the internal energy per unit area given by
\begin{equation}
U = \frac{c_s^2 \Sigma}{\gamma (\gamma - 1)},
\label{intenerg}
\end{equation}
we get \citep{gammie01}
\begin{equation}
\alpha_g = \frac{4}{9 \gamma (\gamma - 1) t_{\rm cool} \Omega}.
\label{alphag}
\end{equation} 

Although we are assuming that the primary transport mechanism is self-gravity, if regions of the disc were hot enough
to be partially ionised, MRI may play a role in these regions.  We therefore assume that if the temperature exceeds
$\sim 1400$ K MRI will operate and set $\alpha = \alpha_{\rm ss} + \alpha_{\rm g} = 0.01$.  It has been suggested \citep{armitage01,zhu09} that this may explain FU Orionis outbursts.  Self-gravity will transport mass to the inner disc where it will pile-up and cause the 
temperature in the inner disc to rise in order to remain gravitationally stable.  Once hot enough for there to be
partial ionisation, MRI will operate, rapidly draining the inner disc producing an FU Orionis-like outburst event.

\subsection{Numerical method}
We consider an initial core mass, $M_c$, with angular velocity, $\Omega_c$ and choose initially a small timestep, $dt$. 
The disc is modelled by a grid with 500 logarithmically spaced grid points extending from $r_{\rm in} = 0.25$ au 
to $r_{\rm out} = 10 a_o$. This does mean that the grid spacing is not the same for different values of $\Omega_c$, but does ensure
that the outer edge of the disc should never reach the outer edge of the grid.  The initial infall is modelled as described in Section
2.1 and we assume that any material that falls inside the inner edge of the disc 
falls directly onto the protostar.  After the first small timestep we then have a small amount of mass on the protostar, and the
rest of the material distributed throughout the disc as described by equation (\ref{shell_frac}).  This means we now know the
disc surface density, $\Sigma$.  The midplane temperature, $T_c$, is then determined using $Q = 2$ with the additional
constraint that $T_c \ge 10$ K. The optical depth, $\tau$, cooling time, $t_{\rm cool}$, and effective
gravitational visocsity, $\nu_g= \alpha_g c_s^2 / \Omega$, can then be determined and the disc is evolved using equation (\ref{visc_evolution}). The amount
of mass accreting onto the star is determined from the difference between the disc mass before and after viscous evolution.  
The new mass distribution in the disc also changes the gravitational potential resulting in a lack of force balance in the
disc.  To achieve force balance material is moved inwards using the method described in \citet{bath81}. The infall is then repeated 
with again some mass falling directly onto the protostar and the rest being distributed
throughout the disc. However, when adding mass to the disc, the added material will generally have a lower specific
angular momentum than the material already in the disc.  This gives rise to an inflow which we take into account using, again,
the method described by \citet{bath81}. This once again changes the gravitational potential and so the lack of force balance
must also be accounted for as before. The new disc surface density can then be used to recalculate the new gravitational
viscosity and the process is repeated. 

After the initial timestep (in which we essentially populate the disc with a small amount of mass), the subsequent 
timesteps are determined after the viscosity has been calculated and are set to be a small fraction of the smallest diffusion
time across a grid cell. We use a zero-torque boundary, $\Sigma(r_{\rm in}) = 0$, at the inner edge of the grid and although the outer
boundary has no influence on the results, we set $v_r$ = 0 at this location.

\section{Results}
We consider core masses from $0.25$ M$_\odot$ to $5$ M$_\odot$.  In all cases we choose the radius of the cloud such that
the resulting uniform density of $\rho = 6 \times 10^{-19}$ g cm$^{-3}$ gives a free-fall time, $t_{\rm ff}$, of $86900$ years. For $M_c = 1$ M$_\odot$ this 
corresponds to a radius of $0.03$ pc. The angular velocity is
parametrised using $f = \Omega_c / \sqrt{G \rho_c}$ and we vary $f$ from $f = 0.05$ to $f = 1.3$, corresponding to
angular velocities between $1.6 \times 10^{-15}$ rad s$^{-1}$ and $2.6 \times 10^{-13}$ rad s$^{-1}$.  In many situations,
the angular velocity is represented as the ratio of the total kinetic energy of rotation to the absolute value of the
gravitational potential energy, often referred to as $\beta$.  The relationship between $\beta$ and $f$ is $\beta = f^2/(2 \pi)$
and so our chosen rotation rates correspond to $\beta$ values between $\beta = 0.0004$ and $\beta = 0.27$. 

\subsection{Quasi-steady evolution}
\begin{figure}
\begin{center}
\psfig{figure=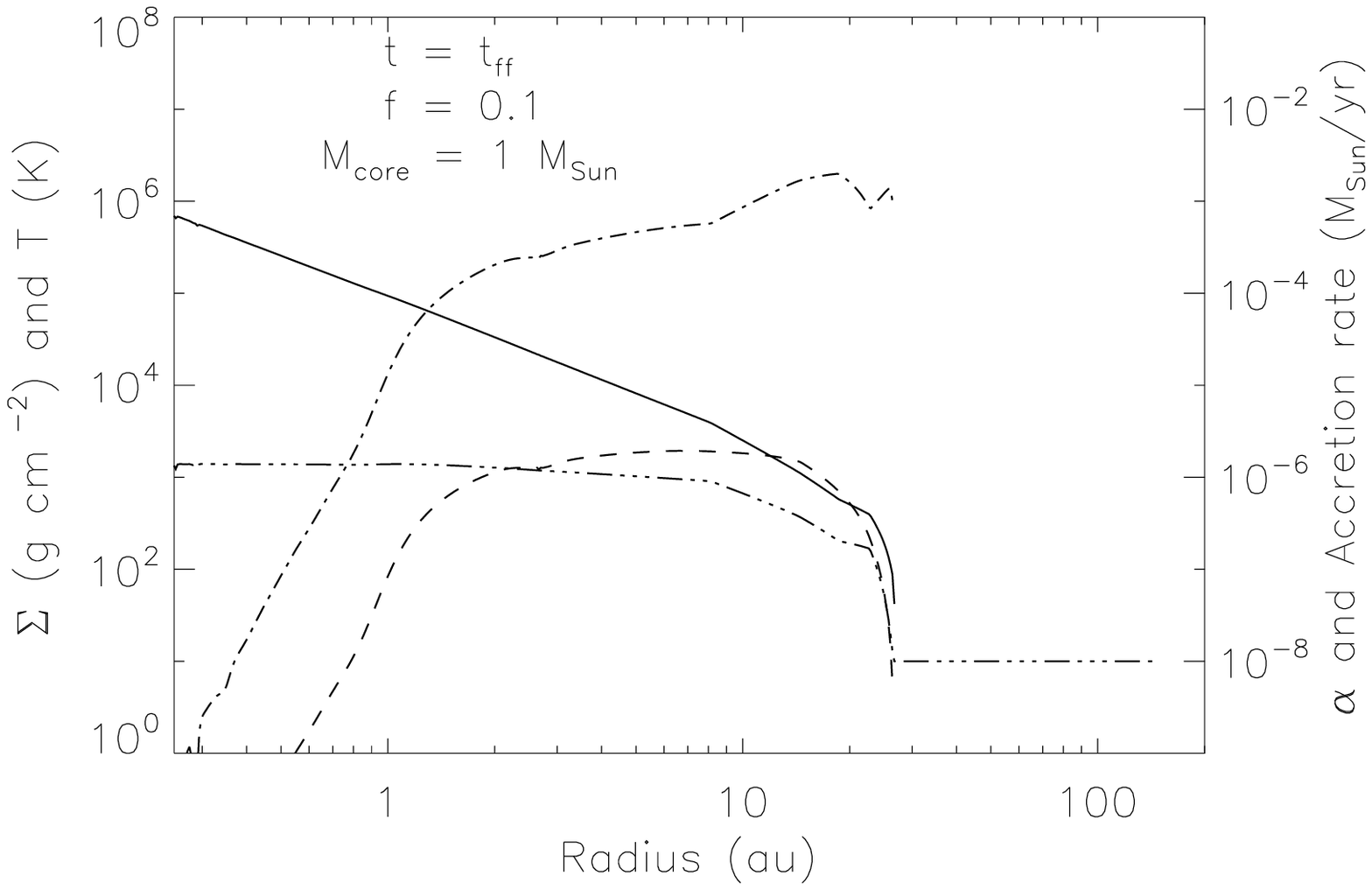,width=0.5\textwidth} 
\psfig{figure=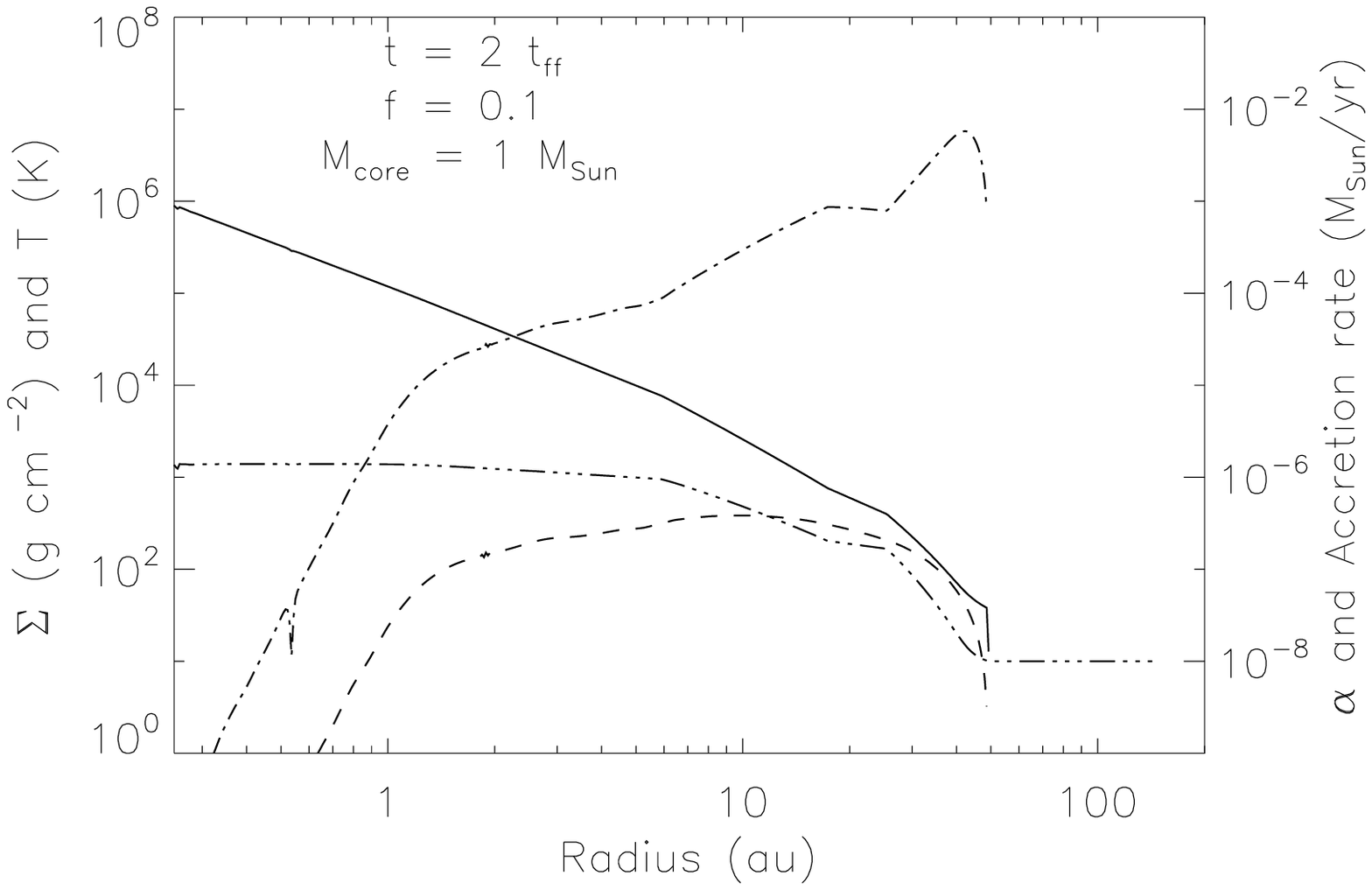,width=0.5\textwidth}
\psfig{figure=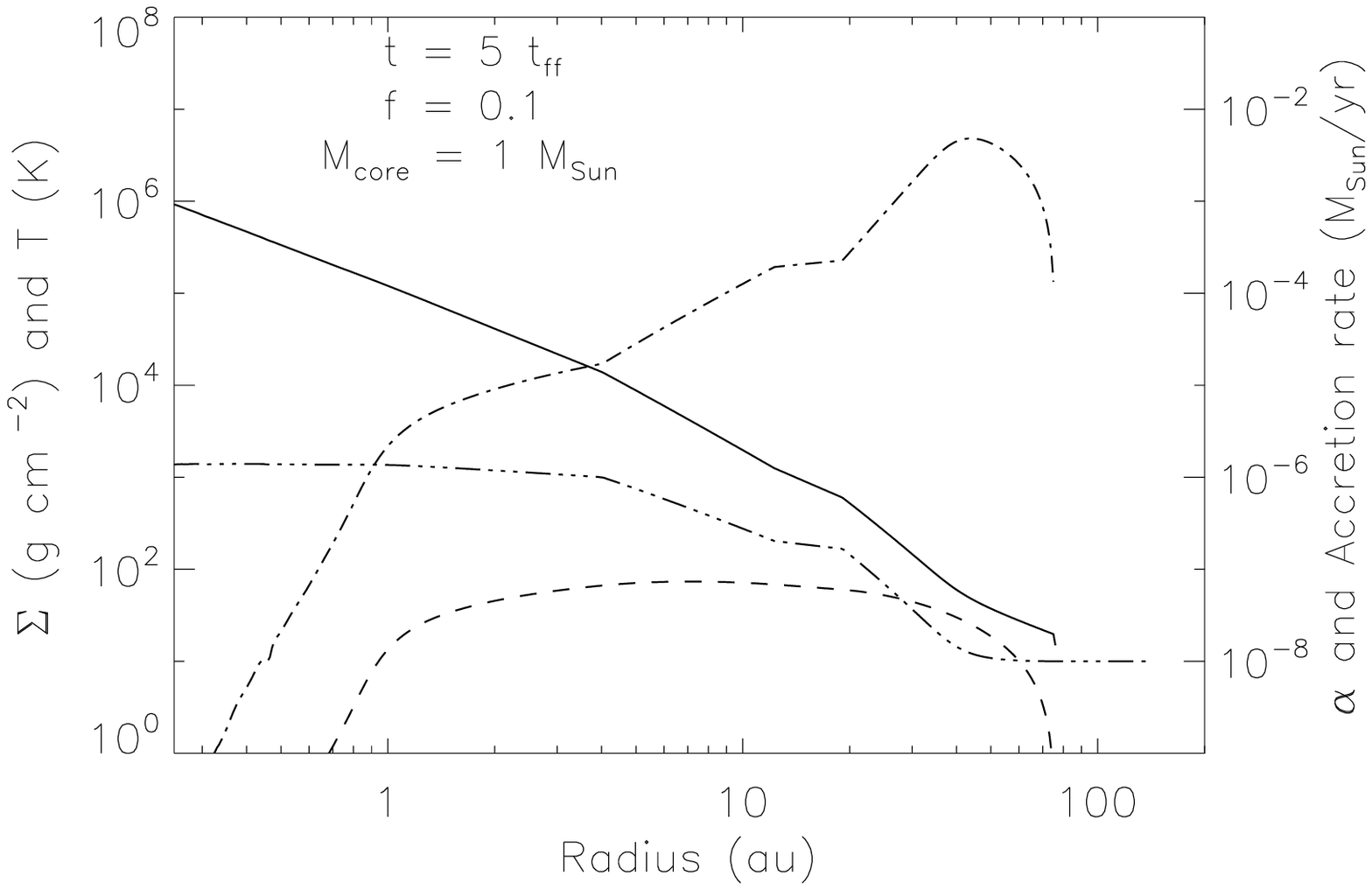,width=0.5\textwidth}
\end{center}
\caption{Disc properties for $M_{\rm core} = 1$ M$_\odot$ for $f = 0.1$  
at $t = t_{\rm ff}$, $t = 2 t_{\rm ff}$, $t = 5 t_{\rm ff}$.  
Each panel shows the surface density $\Sigma$, (solid line), temperature (dash-dot-dot-dot line), effective
gravitational $\alpha$ (dash-dot line), and mass accretion rate (dashed line). The left-hand y-axis
is for $\Sigma$ and temperature, while the right-hand y-axis is for $\alpha$ and mass accretion rate. This
figure illustrates that the disc quickly settles into a quasi-steady state that persists for many free-fall
times.}
\label{quasi-steady_Mc1_f01}
\end{figure}

\begin{figure}
\begin{center}
\psfig{figure=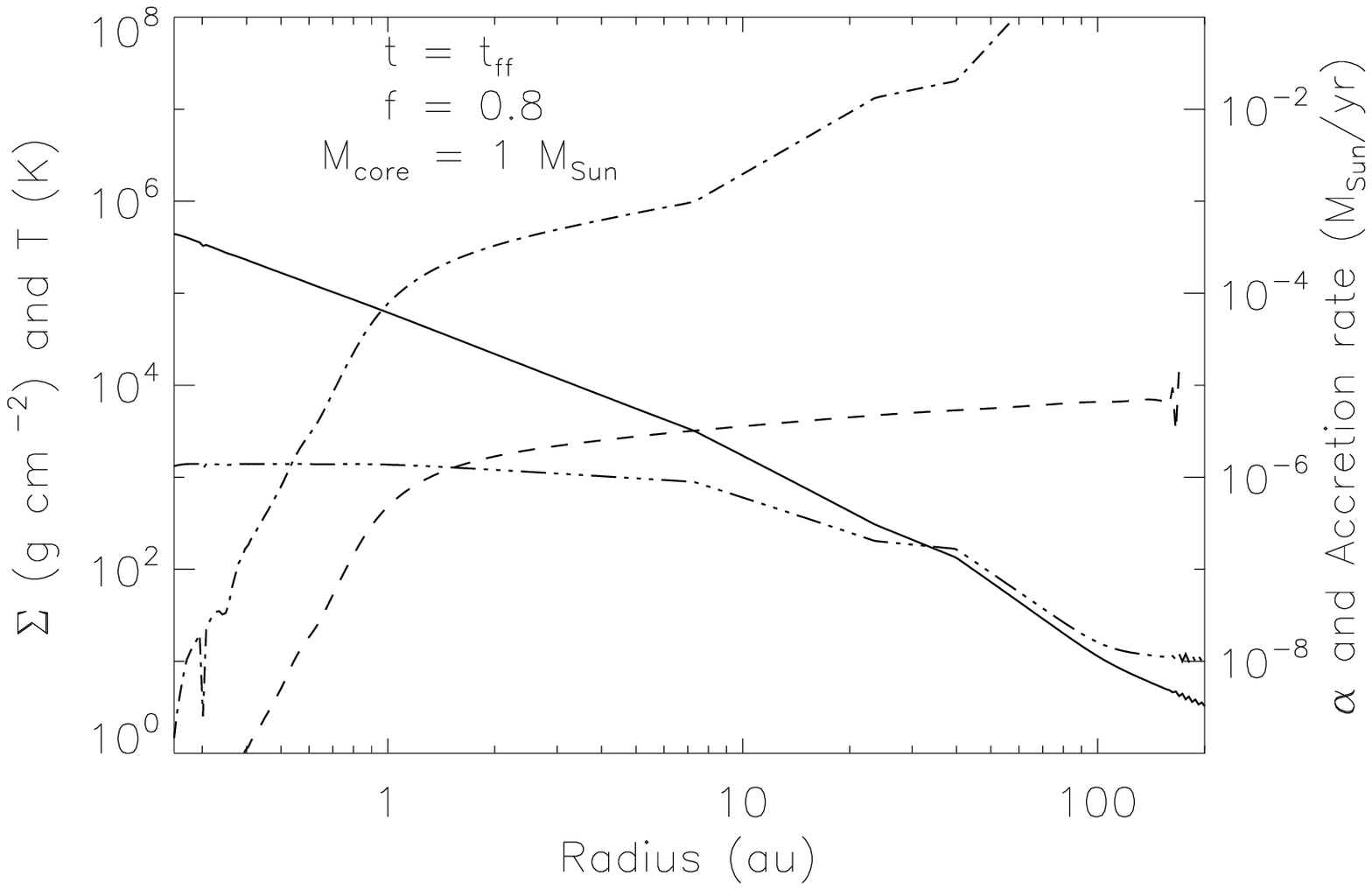,width=0.5\textwidth} 
\psfig{figure=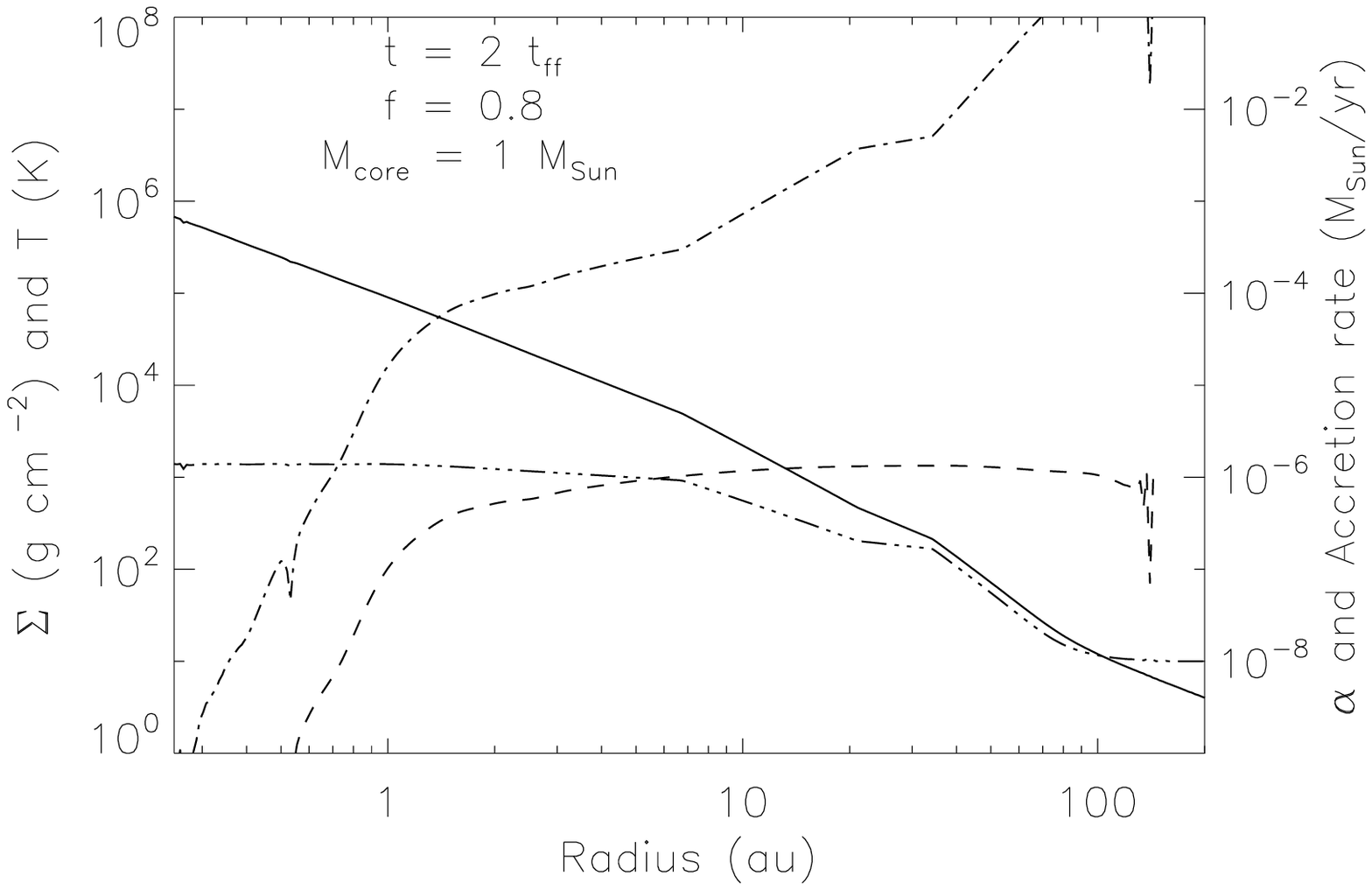,width=0.5\textwidth} 
\psfig{figure=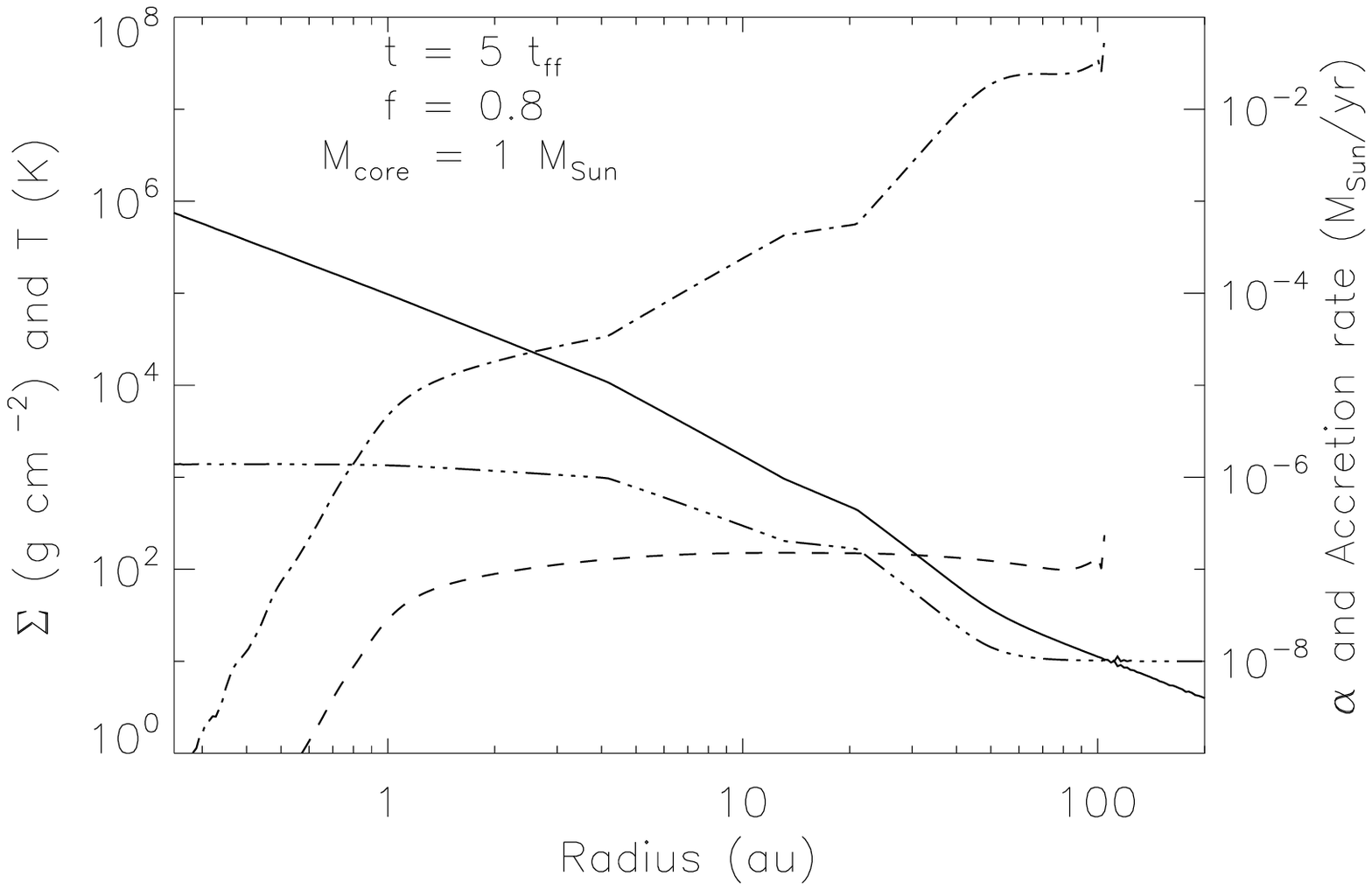,width=0.5\textwidth}
\end{center}
\caption{The same as in Figure \ref{quasi-steady_Mc1_f01} except for $f = 0.8$.  This again shows that
the disc settles into a quasi-steady state that persists for many free-fall times.}
\label{quasi-steady_Mc1_f08}
\end{figure}

It has been suggested \citep{rice09} that a self-gravitating protoplanetary disc will settle into a quasi-steady
state with a mass transfer rate that is approximately the same at all radii.  \citet{rice09}, however, assumed initial power-law surface
density profiles which were evolved until a quasi-steady state was achieved.  Here, we self-consistently
build and evolve the disc through the collapse of a uniform density molecular cloud core. Figure \ref{quasi-steady_Mc1_f01} shows the disc 
properties at $t = t_{\rm ff}$, $t = 2 t_{\rm ff}$, and $t = 5 t_{\rm ff}$ for $M_c = 1$ M$_\odot$ and for $f = 0.1$. 
Each  panel shows the disc surface
density (solid line), central temperature (dash-dot-dot-dot line), effective gravitational $\alpha$ (dash-dot line),
and mass accretion rate (dashed line) all plotted against radius from $0.25$ au to $200$ au.  The left-hand y-axis is scaled for 
surface density and temperature, while the right-hand y-axis is scaled for $\alpha_{\rm g}$ and mass accretion rate.  The mass accretion rate 
is determined using $\dot{M} = 3 \pi \nu \Sigma$ \citep{pringle81}. Figure \ref{quasi-steady_Mc1_f08} is the same as
Figure \ref{quasi-steady_Mc1_f01} except that $f = 0.8$ (i.e., the initial core rotation rate is faster in Figure \ref{quasi-steady_Mc1_f08}
than in Figure \ref{quasi-steady_Mc1_f01}). 

As suggested by \citet{rice09}, the disc quickly settles into a quasi-steady state with a mass transfer
rate that, apart from the very inner disc, is similar at all radii.  The low mass transfer rate in the
inner disc is due to the assumption that MRI operates when $T_c > 1400$ K \citep{zhu09}, which we approximate
by setting $\alpha = \alpha_{\rm ss} + \alpha_{\rm g} = 0.01$ when this condition is satisfied. The disc therefore has short, episodic periods
of rapid accretion (not shown in Figure \ref{quasi-steady_Mc1_f01} or Figure \ref{quasi-steady_Mc1_f08}) draining mass from the inner disc, and
much longer quiescent periods while the inner disc is replenished. We don't, however, claim that we are 
modelling the processes in the inner disc particularly accurately.  A much more detailed analysis has
been carried out by \citet{zhu09b}.

For $f = 0.1$ (Figure \ref{quasi-steady_Mc1_f01}, the disc is quite compact
due to the slow rotation producing a small $a_{\rm max}$. The disc does expand slightly with time, reaching
an outer radius of $\sim 100$ au at $t = 5 t_{\rm ff}$.  For $f = 0.8$ (Figure \ref{quasi-steady_Mc1_f08}) the disc radius is much larger and actually
extends initially beyond the $200$ au shown in Figure \ref{quasi-steady_Mc1_f08}. In both cases, the disc properties are, however,
essentially the same as shown by \citet{rice09} and remain largely unchanged over the 5 free-fall times shown
in both Figure \ref{quasi-steady_Mc1_f01} and Figure \ref{quasi-steady_Mc1_f08}. In this quasi-steady state, the disc has a reasonably steep surface density 
profile ($\Sigma \propto r^{-1.5}$). The $\alpha$ values, on the other hand can increase dramatically with
radius due to the optically thick inner disc resulting in a very long cooling times - and 
hence small $\alpha$ values - and the optically
thin outer disc resulting in short cooling times and large $\alpha$ values. 

The surface density profile is also not a pure power-law, and has a significant fraction of the mass in the inner
disc ($r < 10$ au).  The form of the surface density profile can be largely understood from the temperature
profile.  In the outer disc, the temperature is limited by our assumed minimum of $10$ K.  The temperature rises
with decreasing radius, eventually reaching the temperature ($T_c = 170$ K) above which ice mantles can
no longer exist on the solid grains, generally known as the ice- or snowline.  This produces an abrupt change in the opacity, 
the local cooling time, and consequently the effective gravitational viscosity.  The tendency to evolve
to a state with roughly constant $\dot{M}$ then produces a corresponding change in the surface density profile. The temperature 
then continues to rise with decreasing radius, eventually reaching $1400$ K and becoming sufficiently ionised for
MRI to operate. This sets an effective maximum temperature and also produces a subtle change in the surface density profile.
A detailed description of how the surface density and temperarture profiles vary in the different opacity
regimes is included in \citet{clarke09}. 

\begin{figure}
\begin{center}
\psfig{figure=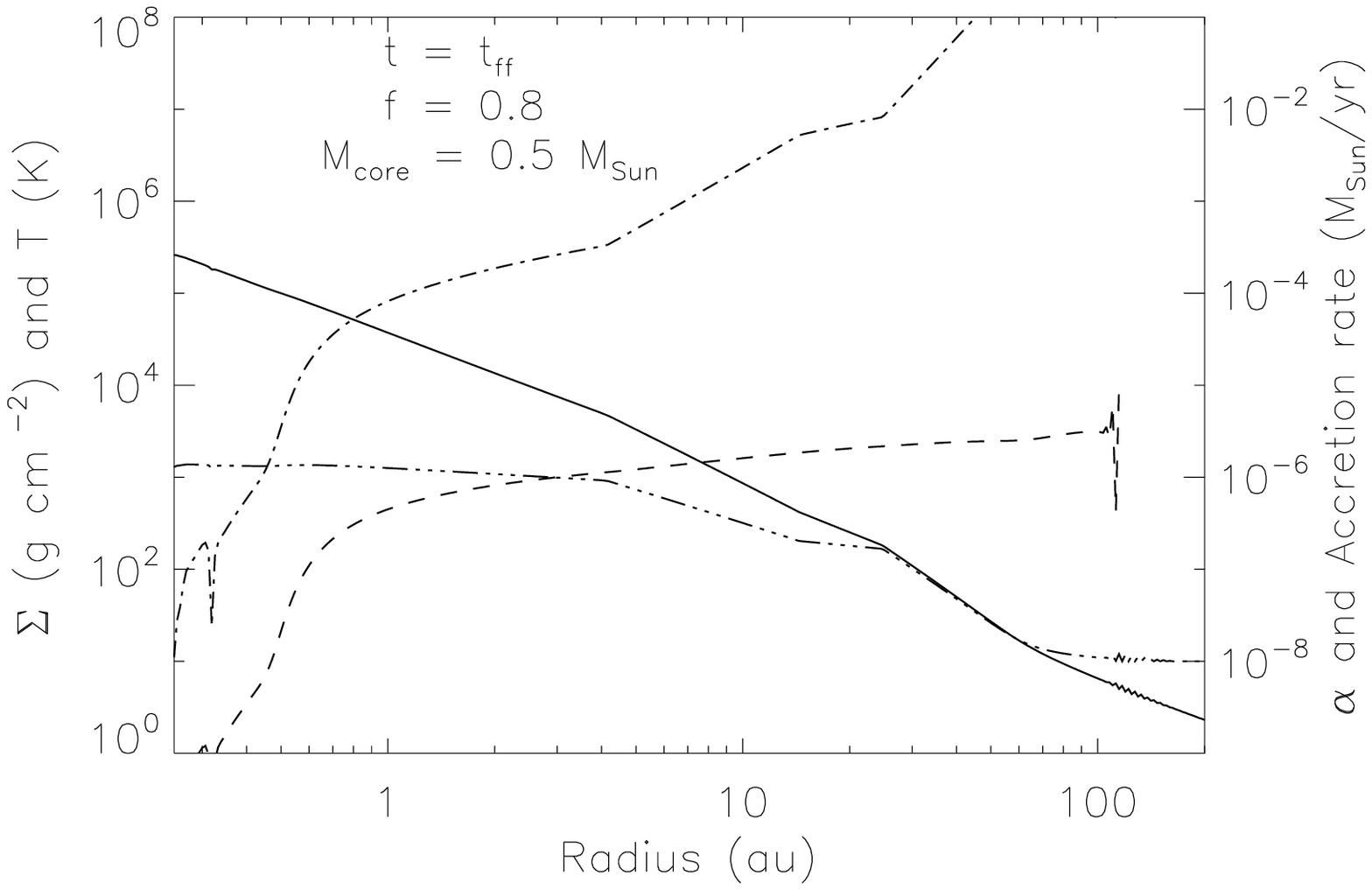,width=0.5\textwidth} 
\psfig{figure=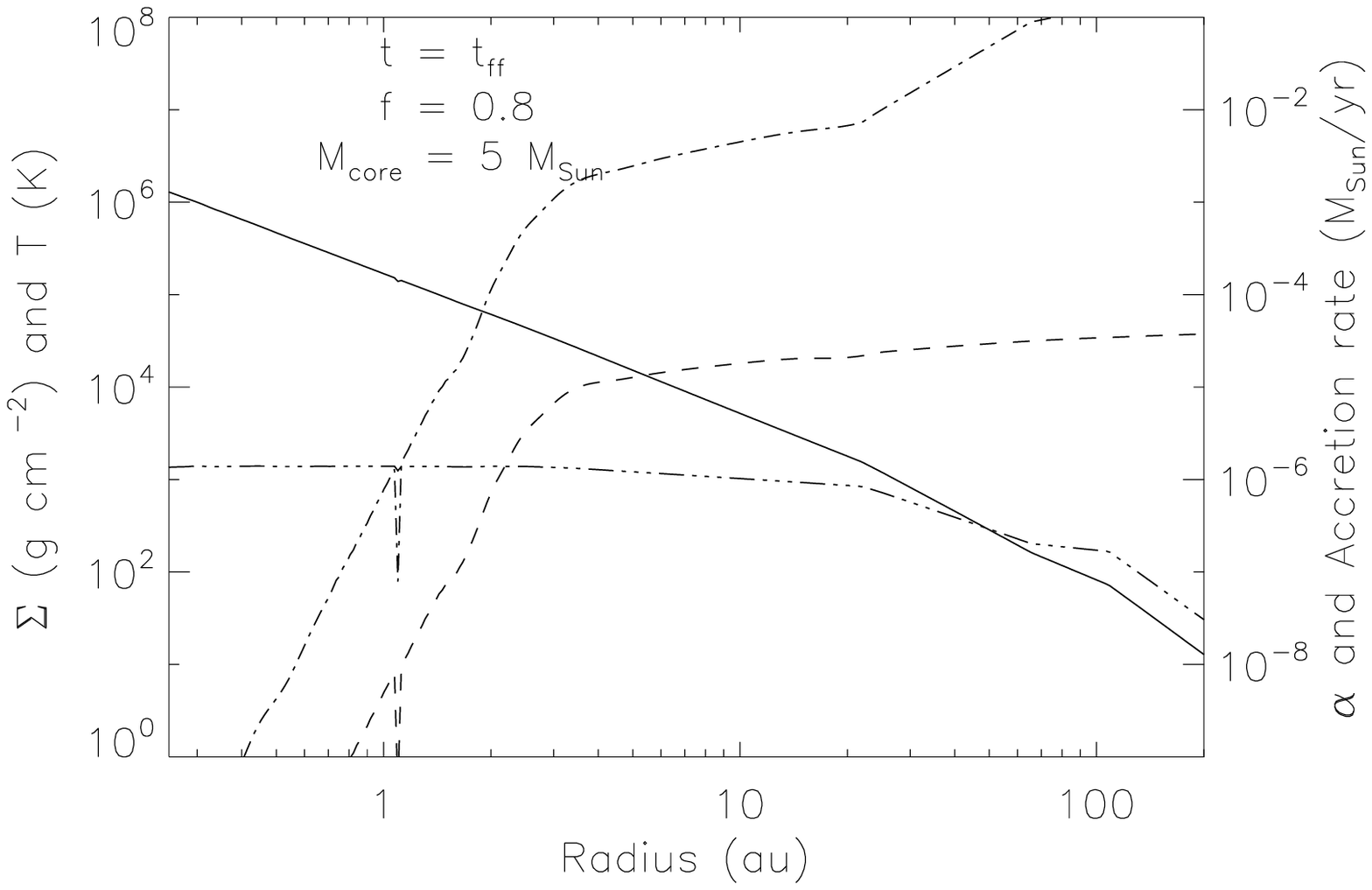,width=0.5\textwidth} 
\end{center}
\caption{Disc properties for $f = 0.8$ and for $M_{\rm core} = 0.5$ M$_\odot$ (top panel) and $M_{\rm core} = 5$ M$_\odot$.
This figure illustrates that the quasi-steady nature of self-gravitating discs is largely the same for all
core masses. There are some quantitative differences such as the surface density, the radius of the ice/snowline,
and the radial extent over which MRI can operate all increase with increasing core mass.}
\label{Mc05Mc5}
\end{figure}

Not only does the quasi-steady nature of the disc persist for a number of free-fall times, it is also qualitatively
the same for all our chosen parameters.  Figure \ref{Mc05Mc5} shows the disc properties (as in
Figure \ref{quasi-steady_Mc1_f01} and Figure \ref{quasi-steady_Mc1_f08}) at $t = t_{\rm ff}$ for $f = 0.8$ and for $M_c = 0.5$ M$_\odot$
(top panel) and 
$M_c = 5$ M$_\odot$ (bottom panel). The form of the profiles is essentially the same as in Figures \ref{quasi-steady_Mc1_f01} and 
\ref{quasi-steady_Mc1_f08}. A few quantitative
differences are that the surface density, the radius of the ice/snowline, and the radial extent of the region over which MRI
can operate all increase with increasing core mass.
Figure \ref{quasi-steady_Mc1_f08} 
also illustrates how the snowline moves in with time starting at $\sim 50$ au when
$t = t_{\rm ff}$ and moving in towards $10$ au at $t = 5 t_{\rm ff}$.   

\subsection{Mass accretion rates}
\begin{figure}
\begin{center}
\psfig{figure=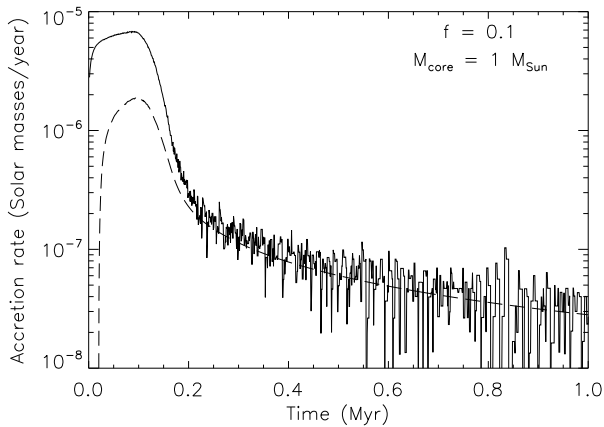,width=0.5\textwidth} 
\psfig{figure=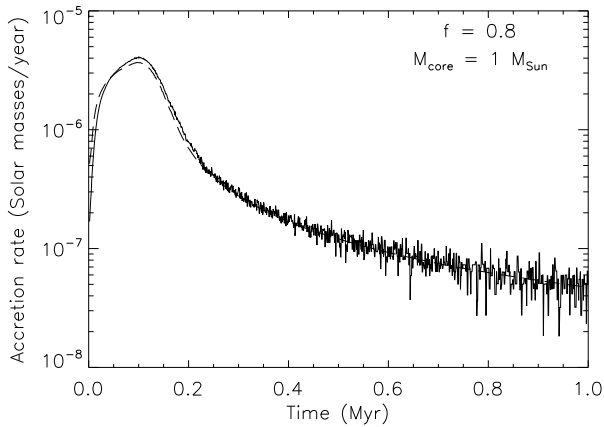,width=0.5\textwidth} 
\end{center}
\caption{Figure showing the accretion rate onto the protostar (solid line) and through the disc
(dashed line) for $M_{\rm core} = 1$ M$_\odot$ and $f = 0.1$ (top panel) and $f = 0.8$ (bottom panel).
The accretion rate initially increases with time, peaking at $t \sim t_{\rm ff}$.
The accretion onto the protostar is also quite variable and in fact has been averaged 
over $0.1 t_{\rm ff}$ so is
actually more variable than shown.  The disc accretion is, however, much steadier and
illustrates the quasi-steady nature of the disc evolution.  
The figure also shows that for high core rotation rates ($f = 0.8$) accretion onto the protostar 
is governed by accretion through the disc, while for slow core rotation rates ($f = 0.1$) the
protostar grows primarily through direct infall during the first 2 free-fall times.}
\label{Mdot_Mc1}
\end{figure}

We compute the accretion rate onto the protostar by simply determining how its mass changes with time. The disc
accretion rate is computed using $\dot{M} = 3 \pi \nu \Sigma$ measured at a fixed location in the disc 
(generally $3$ au for slowly rotating cores and $10$ au for rapidly rotating cores). Figure \ref{Mdot_Mc1} shows how the disc accretion
rate (dashed-line) and protostar accretion rate vary with time for $M_c = 1$ M$_\odot$ and for $f = 0.1$ 
(top panel) and $f = 0.8$ (bottom panel).  In both cases shown the protostar accretion rate rises rapidly
with time and peaks at $t \sim t_{\rm ff}$ ($\sim 10^5$ years).  It then drops significantly over the next freefall 
time and then changes more slowly from $t = 2 t_{\rm ff}$ to $t = 1$ Myr. The protostar accretion rate also becomes 
quite variable after $t = 2 t_{\rm ff}$ due to the accretion onto the protostar occuring episodically when $T_c > 1400$ K.
In fact the protostar accretion rate has been averged over $0.1 t_{\rm ff}$ and so the variability that our model produces
is somewhat
more extreme than shown in Figure \ref{Mdot_Mc1}, although we don't claim to be modelling this variability particularly
accurately.  

The disc accretion rate (dashed line) is much smoother than the protostar accretion rate and, in Figure \ref{Mdot_Mc1}, has not been averaged.
This illustrates that the disc is able to reach a quasi-steady state on timescales much shorter than the freefall time.
What Figure \ref{Mdot_Mc1} also shows is that for rapidly rotating cores (large $f$) the accretion onto the core is governed
by accretion through the disc, while for slowly rotating cores the protostar grows initially through
direct infall and only after infall ceases ($t > 2 t_{\rm ff}$) does disc accretion determine the rate at which mass is
accreted onto the protostar. The accretion rate at $t = 1$ Myr also depends on the core rotation and increases with
increasing $f$. From Figures \ref{quasi-steady_Mc1_f01}, \ref{quasi-steady_Mc1_f08} and \ref{Mc05Mc5}, it is clear that the surface density profiles do not vary significantly
with $f$ or with time. What is different is the mass of the central protostar. The protostar mass increases with decreasing
$f$ and consequently (since we assume $Q = 2$) produces disc temperatures that also decrease with decreasing $f$.
This means that the cooling rate, effective gravitational viscosity, and consequently the disc accretion rate all decrease
with decreasing $f$. In fact for very slow rotating cores ($f < 0.08$) it is difficult to sustain disc accretion beyond a
few freefall times. 

\begin{figure}
\begin{center}
\psfig{figure=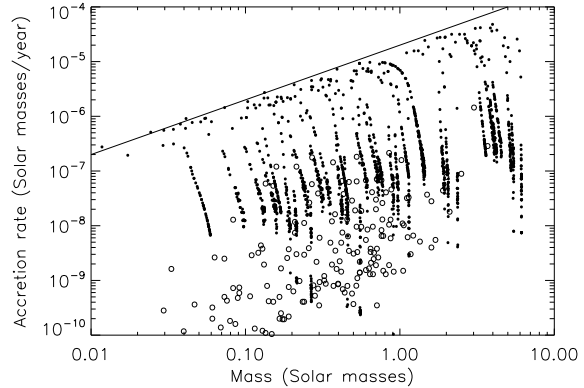,width=0.45\textwidth}
\end{center}
\caption{Protostar accretions rates plotted against protostar mass
from all our simulations (solid dots). The data points are at $t = t_{\rm ff}$ and at 50
random chosen times between $t = t_{\rm ff}$ and 1 Myr.  The sparsely populated region
at the top illustrates how the accretion rate drops rapidly between $t = t_{\rm ff}$ and
$t = 2 t_{\rm ff}$ and suggests that we are not particularly likely to observe systems in
this state.  The open circles show observed accretion rates for TTauri stars taken
from \citet{gullbring98}, \citet{white01}, \citet{calvet04}, \citet{natta06}. Although
these don't compare particularly well with our calculated accretion rates, the highest
observed accretion rates are consistent with what we would expect for the youngest systems, and
many of the observed systems are probably at a later stage of evolution than the
simulated systems. The solid line shows that the peak of the simulated accretion rates
depend linearly on protostar mass ($\dot{M} \propto M$) and has the same relation as the upper
envelope of the observed accretion rates \citep{hartmann06}.}
\label{Mdot_all}
\end{figure}

Since we have carried out a large number of simulations with various core masses and with various rotation rates, we
can consider how the mass accretion rate varies with protostar mass during the first $1$ Myr. Figure \ref{Mdot_all}
shows the mass accretion rate against protostar mass from our simulations (solid dots) compared with observed
accretion rates for T Tauri stars taken from \citet{gullbring98}, \citet{white01}, \citet{calvet04}, and \citet{natta06}. 
The data points from our simulations are at $t = t_{\rm ff}$ (the peak accretion rate) and then at 50 randomly
chosen times between $t = t_{\rm ff}$ and
$t = 1$ Myr.  The reasonably sparsely populated upper region illustrates how the accretion rate drops, within
about a freefall time, to a value an order of magnitude or more lower than the peak value. The lower region is more densely
populated, making it more likely that observed accretion rates would fall in this band.  This likelihood is increased
further due to the system being more heavily embedded during the first few freefall times, making it harder to measure 
accurate accretion rates at these early times.

The accretion rates from the simulations don't compare particularly well with the observed
accretion rates (open circles), with the observed accretion rates tending to be lower than the 
those from the simulations. The observed systems are, however, 
probably at a later stage of evolution than the simulated systems. The highest
observed accretion rates are, however, consistent with what we would expect for the
youngest systems, suggesting that self-gravitating transport can dominate until the earliest
stages of the TTauri phase. It has also been suggested that the accretion
rate should vary approximately with mass as $\dot{M} \propto M^{2}$ \citep{muzerolle05, natta06}.
The upper boundary of our simulated accretion rates, however, has a mass dependence closer to 
$\dot{M} \propto M$ (illustrated by the solid line in Figure \ref{Mdot_all}) 
and is consistent with that suggested by \citet{vorobyov09} and with the upper
envelope  of observed accretion rates \citep{hartmann06}. That the observed accretion rates
are for protostars of various ages will introduce a scatter that could result in what
appears to be a steeper dependence on protostar mass than is actually the case. 
   
\subsection{Disc fragmentation}
It has been proposed \citep{boss98,boss02} that discs that are sufficiently gravitationally unstable
could fragment to produce bound objects and that these bound objects could subsequently contract to form
either gas giant planets or brown dwarf stars.  What is now well known is that disc fragmentation
requires rapid cooling \citep{gammie01,rice03} and that there is a minimum cooling time - which may depend
on the actual cooling function \citep{cossins09} - for which a
self-gravitating disc can remain in a quasi-steady state without fragmenting.  The relationship
between effective gravitational $\alpha_{\rm g}$ and cooling time (see equation (\ref{alphag})) allows us to
define the fragmentation boundary in terms of $\alpha_{\rm g}$. It has been shown \citep{rice05} that 
fragmentation occurs for $\alpha_{\rm g} > 0.06$ and that this boundary is independent of the specific
heat ratio $\gamma$.  

\begin{figure}
\begin{center}
\psfig{figure=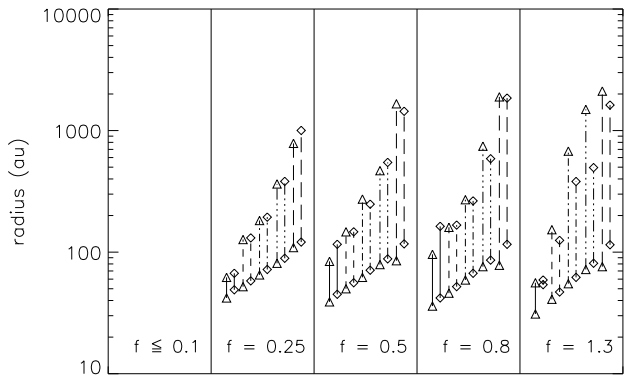,width=0.5\textwidth}
\end{center}
\caption{Figure showing the range of radii where fragmentation could occur 
(defined as the region where $\alpha > 0.1$). We find that the fragmentation
conditions are never satisfied when $f \le 0.1$ and are always satisfied
when $f \ge 0.25$. 
To illustrate that the fragmentation
conditions can be satisfied for many dynamical times, the fragmentation range is plotted at $t = t_{\rm ff}$
(triangles) and $t = 1.5 t_{\rm ff}$ (diamonds).  The core masses considered are $M_{\rm core} = 0.25$
M$_\odot$ (solid line), $M_{\rm core} = 0.5$ M$_\odot$ (dashed line), $M_{\rm core} = 1$ M$_\odot$
(dash-dot line), $M_{\rm core} = 2$ M$_\odot$ (dash-dot-dot-dot line), and $M_{\rm core} = 5$ M$_\odot$
(long-dash line).  The inner fragmentation radius increases with core mass from $\sim 40$ au for 
$M_{\rm core} = 0.25$ M$_\odot$ to $\sim 100$ au for $M_{\rm core} = 5$ M$_\odot$. The radial range also increases
with increasing core mass, but this should be interpreted cautiously as the outer disc could be susceptible
to ionisation from cosmic rays that may stimulate MRI turbulence \citep{rafikov09} 
and stabilise these regions against fragmentation.}
\label{frag_rad}
\end{figure}

Figures \ref{quasi-steady_Mc1_f01}, \ref{quasi-steady_Mc1_f08}, and \ref{Mc05Mc5} show that in all cases, the $\alpha_{\rm g}$ values
in the inner disc are well below that required for fragmentation ($\alpha_{\rm g} < 10^{-2}$), but that the outer parts of
some discs could be susceptible to fragmentation. The requirement for fragmentation is that $Q \sim 1$ and
$\alpha_{\rm g} > 0.06$.  In our simulations we assume $Q$ settles to $Q = 2$, so we don't quite satisfy the first
condition, but we can at least identify the regions where the $\alpha_{\rm g}$ condition is satisified and therefore
where $Q$ will decrease and fragmentation is most likely. 
Figure \ref{frag_rad} shows the range of radii 
where fragmentation could occur at $t = t_{\rm ff}$ (triangles) and $t = 1.5 t_{\rm ff}$ (diamonds),
for various $f$ values.
Each vertical line is for a single core mass and we consider core masses of $M_c = 0.25$ M$_\odot$ (solid line),
$M_c = 0.5$ M$_\odot$ (dashed line), $M_c = 1$ M$_\odot$ (dash-dot line), $M_c = 2$ M$_\odot$ (dash-dot-dot-dot line),
and $M_c = 5$ M$_\odot$ (long dash line).

Figure \ref{frag_rad} shows that only the outer disc (beyond $\sim 40$ au) has conditions suitable for fragmentation,
consistent with numerical simulations \citep{stamatellos07} and possibly also with some observations \citep{greaves08}.
The inner radius of the fragmentation zone increases with core mass and for a core mass of $1$ M$_\odot$,
which produces a central protostar with a mass close to $1$ M$_\odot$, is 
$60 - 70$ au, consistent with \citet{clarke09}.  The radial range that can undergo fragmentation is quite
small for low core masses ($10 - 20$ au for $M_c = 0.25$ M$_\odot$) but increases
dramatically with core mass.  This result needs to be interpreted somewhat cautiously as the 
outer parts of these discs could be susceptible to ionization through cosmic rays that may stimulate
MRI turbulence \citep{rafikov09} and stabilise these regions against fragmentation.  

Although we can't really predict the possible masses of the objects that may form via fragmentation,
\citet{kratter09} suggest that the initial fragmentation masses will be $\sim 1$ M$_{\rm Jupiter}$. 
The large amount of mass in these outer regions, however, suggests that such fragments will continue
to grow and that the isolation mass can easily be $> 100$ M$_{\rm Jupiter}$.  Fragmentation in the outer
parts of these discs is therefore unlikely to form planetary mass objects and will probably
form sub-stellar or stellar mass companions.   

None of our simulations with $f = 0.1$ or less had any regions susceptible to fragmentation.  This was essentially
due to these slow rotation rates producing compact discs with most of the mass in the optically thick inner disc
and very little mass in the optically thin outer disc.  It has been suggested \citep{miyama84} that fragmentation
depends on both the ratio of the thermal to gravitational energy $\xi$ and the ratio of the energy of rotation
to the gravitational energy $\beta = f^2/(2 \pi)$. \citet{miyama84} find that fragmentation occurs if 
$\xi \beta < 0.12$.  If we assume an initially isothermal core, with sound speed $c_s$, then (see \citet{walch09})
\begin{equation}
\xi = \frac{2 c_s^2 R_c}{G M_c}.
\end{equation}   
Since we assume all our cores have the same density and hence freefall time, the $\xi$ values of our cores vary
with core mass.  Since fragmentation is possible for all core masses with $f \ge 0.25$ and for
none of the core masses when $f \le 0.1$, this
suggests that what primarily determines if a disc will fragment is the initial rotation rate of the cloud core. The
reason for this is that, in our simulations, the discs very quickly settle into a quasi-steady state, with $Q = 2$, that is in thermal
equilibrium. This quasi-steady state is essentially independent of the initial thermal properties of the core and depends
largely on how much mass is in the system.  If the initial rotation is such that some of this mass can be deposited
in the outer, optically thin parts of the disc, then fragmentation is quite likely. Figure \ref{frag_rad} also
shows that the fragmentation region remains largely unchanged for at least $0.5 t_{\rm ff}$ (from $t = t_{\rm ff}$ to
$t = 1.5 t_{\rm ff}$). We find, in fact, that conditions
suitable for fragmentation actually persist for a freefall time or longer (which is many dynamical times even in the outer disc), 
but are generally no longer satisfied after $\sim 2$ freefall times for the lower mass cores and after 
$3 - 4$ freefall times for the higher mass cores.   

\subsection{Disc and protostar masses}
\begin{figure}
\begin{center}
\psfig{figure=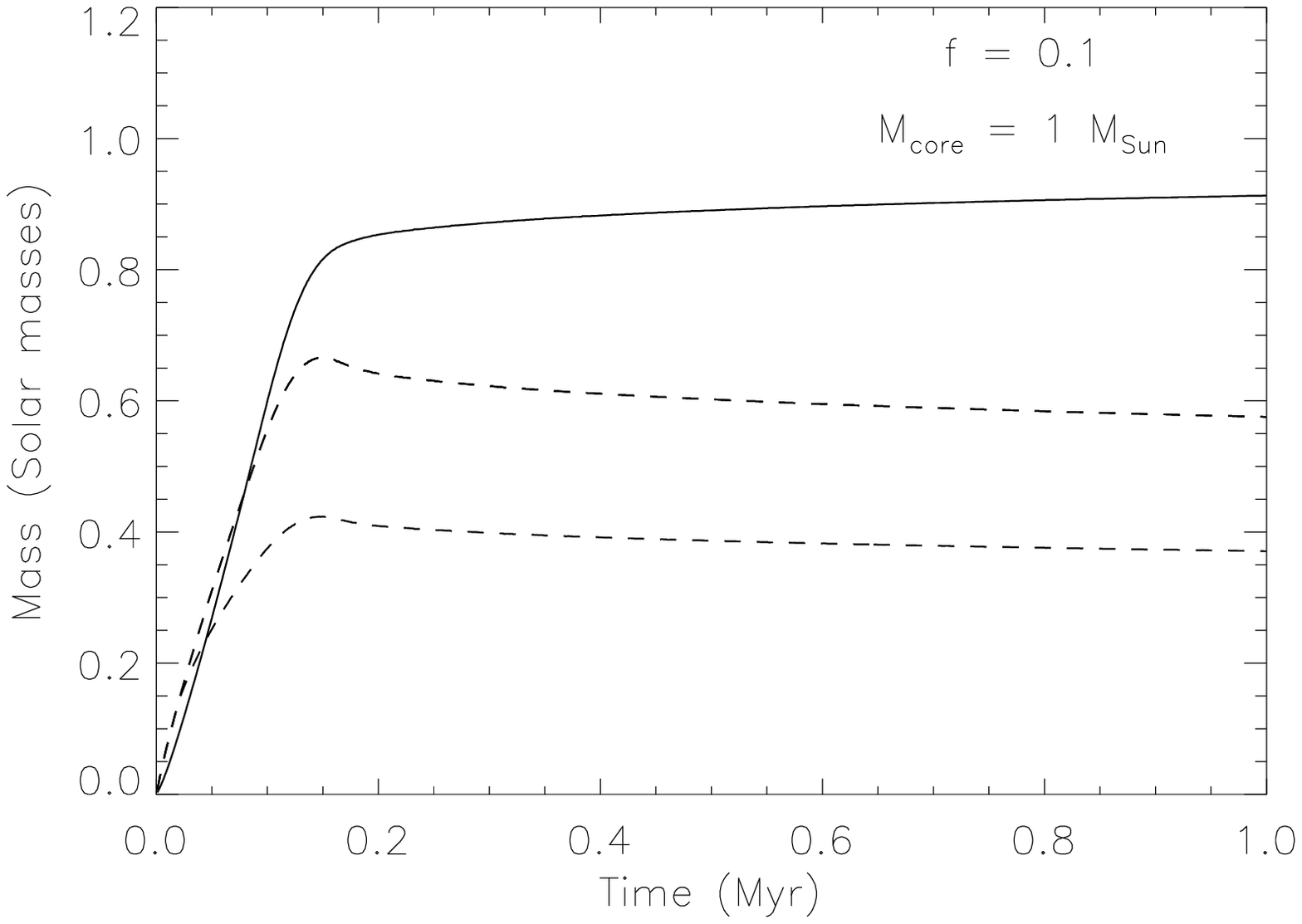,width=0.5\textwidth} 
\psfig{figure=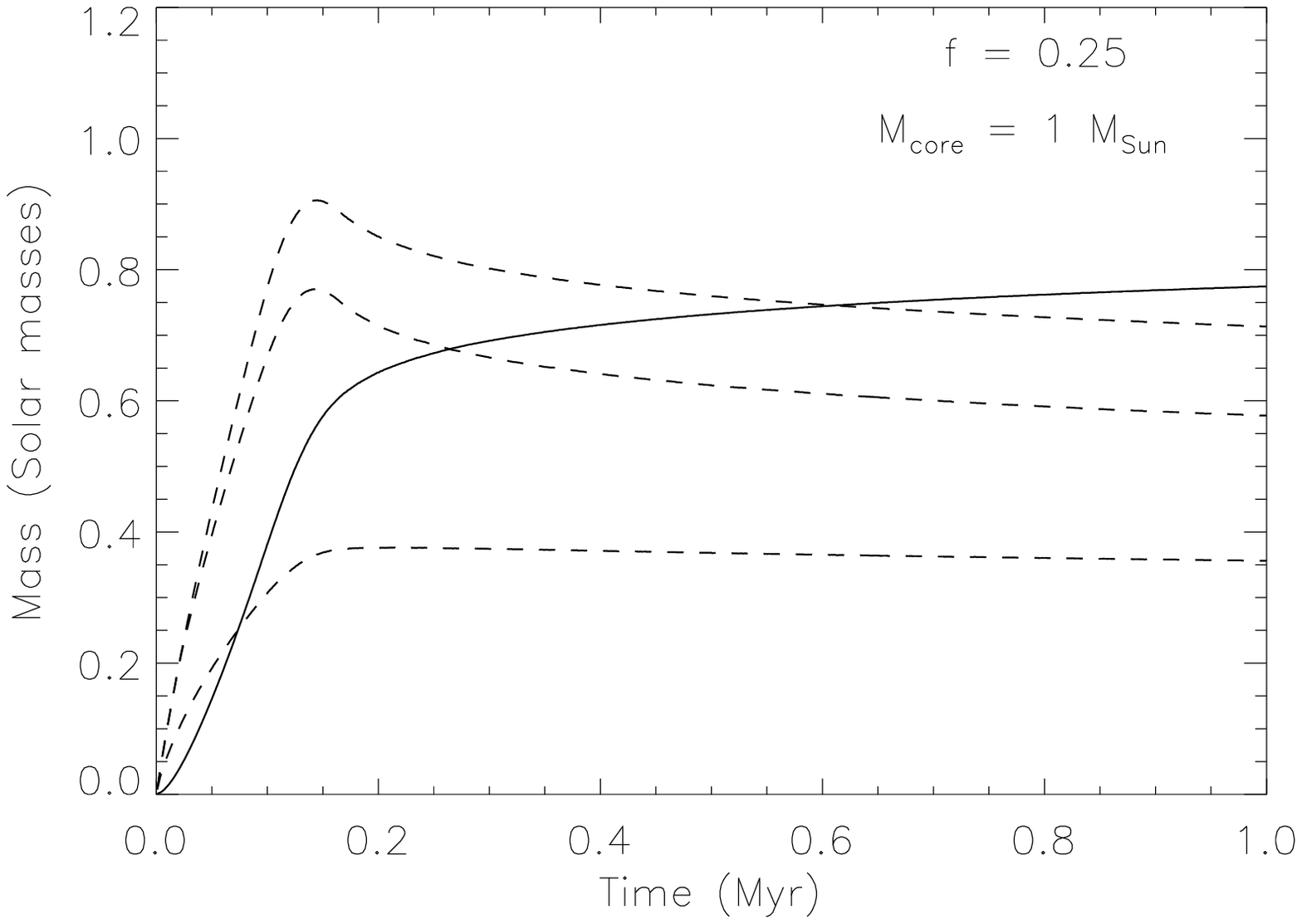,width=0.5\textwidth}
\psfig{figure=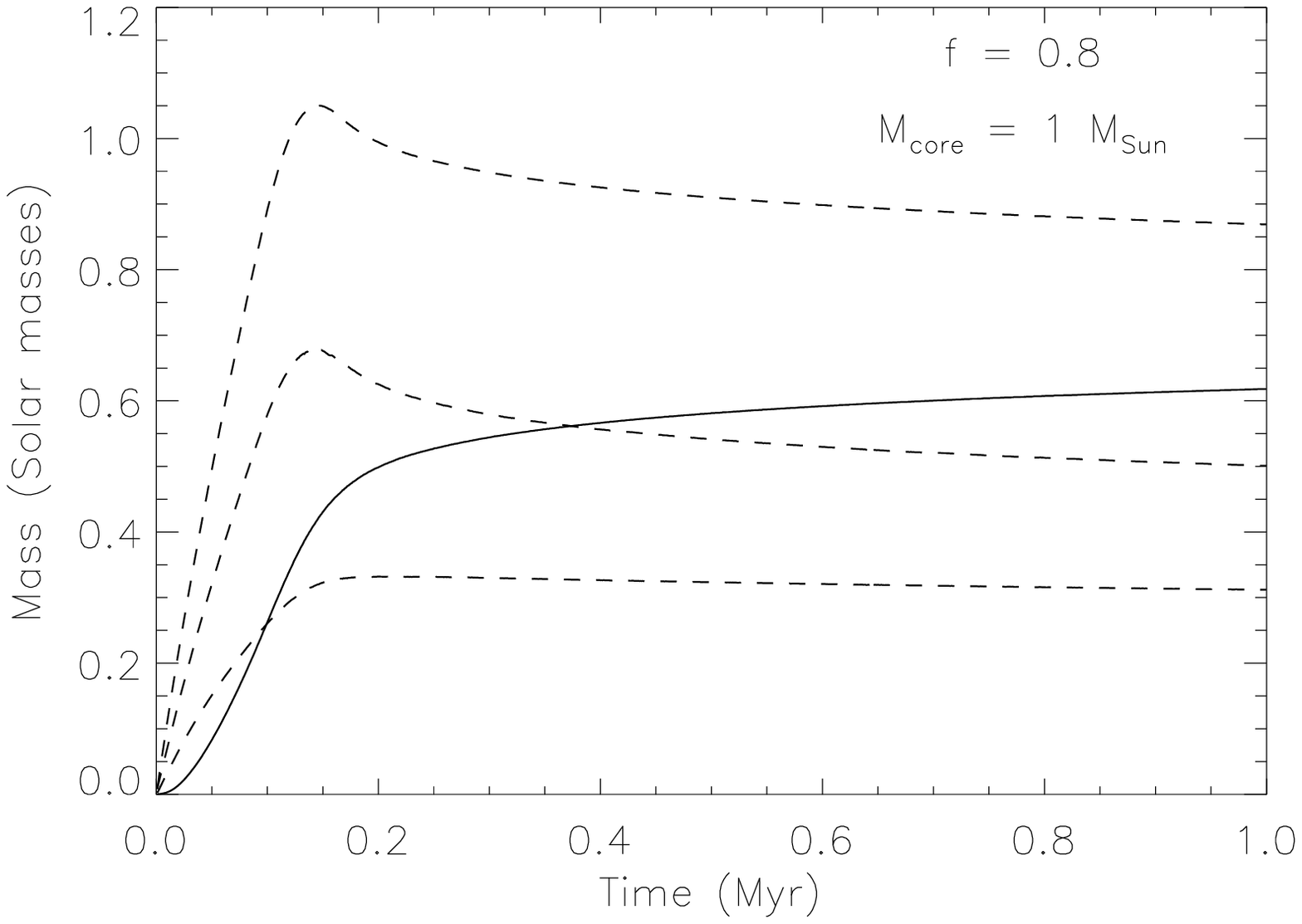,width=0.5\textwidth}
\end{center}
\caption{Protostar (solid line) and disc (dashed lines) masses plotted against time for $M_{\rm core} = 1$ M$_\odot$
and for $f = 0.1$ (top panel), $f = 0.25$ (middle panel), and $f = 0.8$ (bottom panel). The different
dashed lines show the disc mass inside $10$ au, $100$ au, and the total disc mass. The figure illustrates
that as the core rotation rate increases, the core mass decreases and more and more mass is deposited 
at large radii in the disc. 
The figure shows that in all cases a significant fraction ($\sim 50$ \%) of the mass inside $100$ au is located in
the optically thick inner $10$ au. 
}
\label{Disc_protostar_Mc1}
\end{figure}

Figure \ref{Disc_protostar_Mc1} shows the evolution of the disc and protostar masses against time for $M_{\rm core} =
 1$ M$_\odot$ and for $f = 0.1$ (top panel), $f = 0.25$ (middle panel), and $f = 0.8$ (bottom panel).
In all cases, the solid line shows the evolution of the core mass, while the dashed lines show the
disc mass within 10 au, 100 au, and the total disc mass.  For $f = 0.1$ the disc does not extend beyond 
$100$ au and so the mass inside $100$ au and the total disc mass are the same. The figure illustrates that as the core
rotation rate increases, the core mass decreases and more and more mass is deposited at large radii ($r > 100$ au). 
What these figures also illustrate is that the mass inside $10$ au and the mass inside $100$ is only weakly dependant
on the rotation rate, again illustrating the quasi-steady nature of these discs. 

\begin{figure}
\begin{center}
\psfig{figure=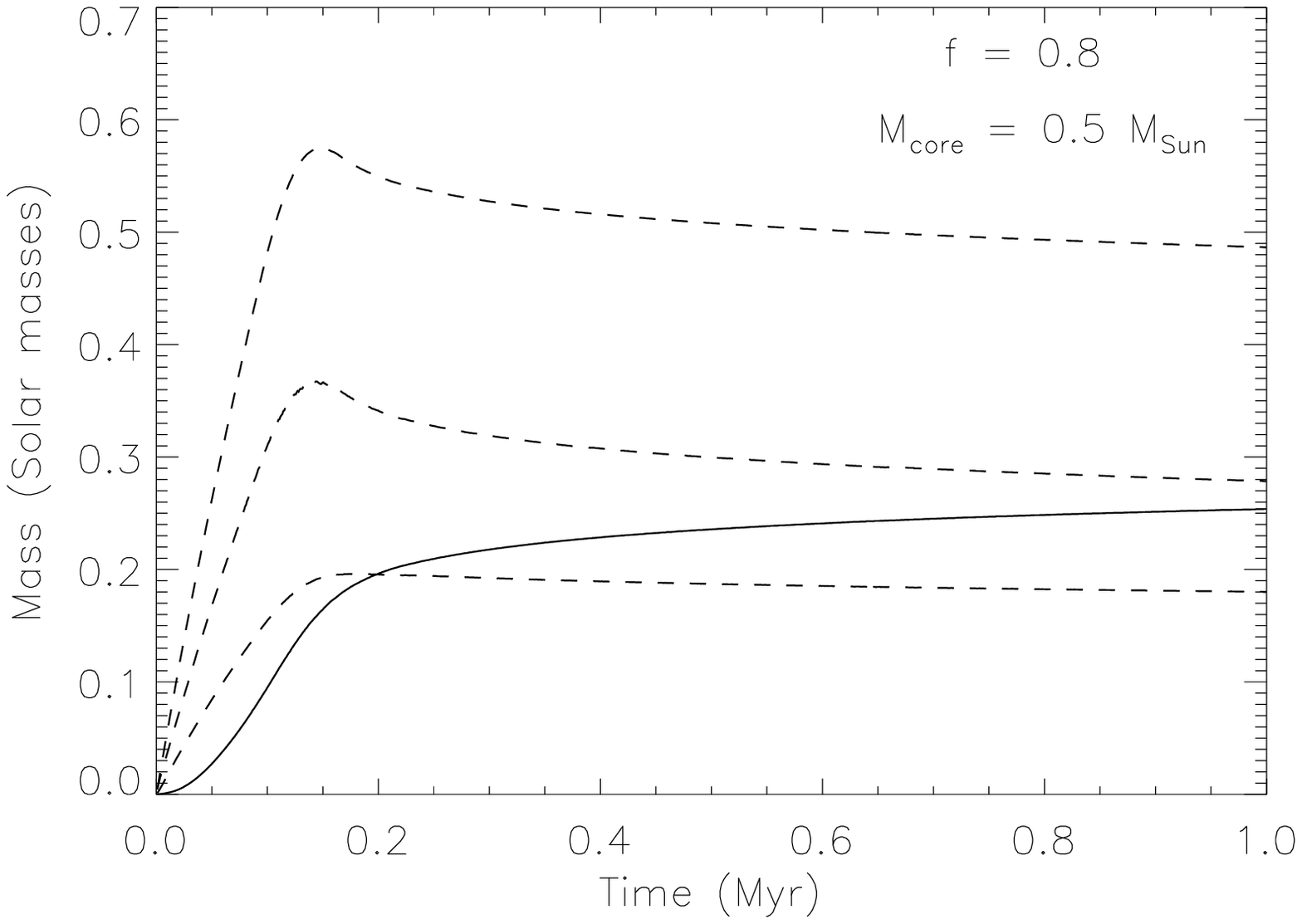,width=0.5\textwidth} 
\psfig{figure=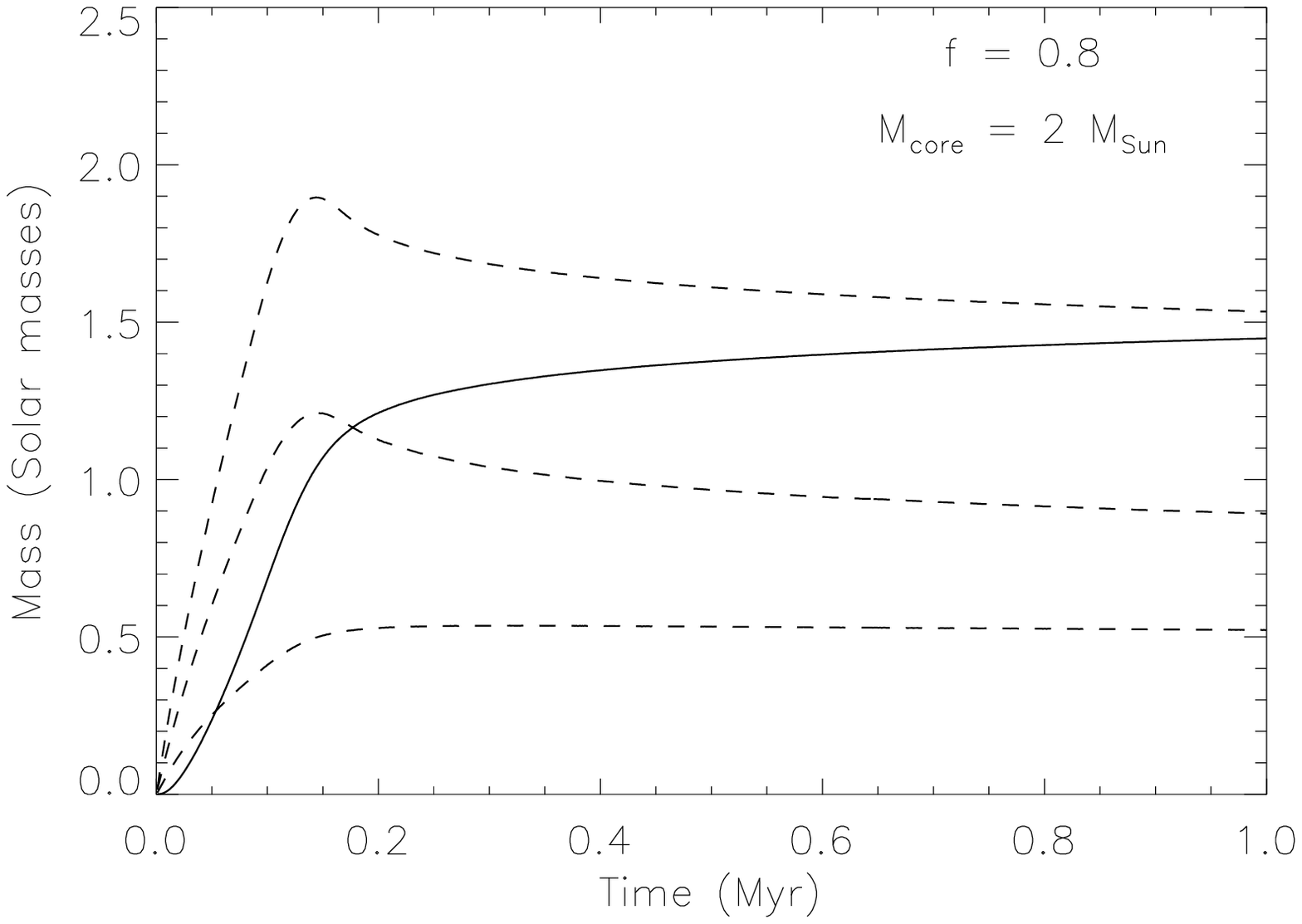,width=0.5\textwidth}
\psfig{figure=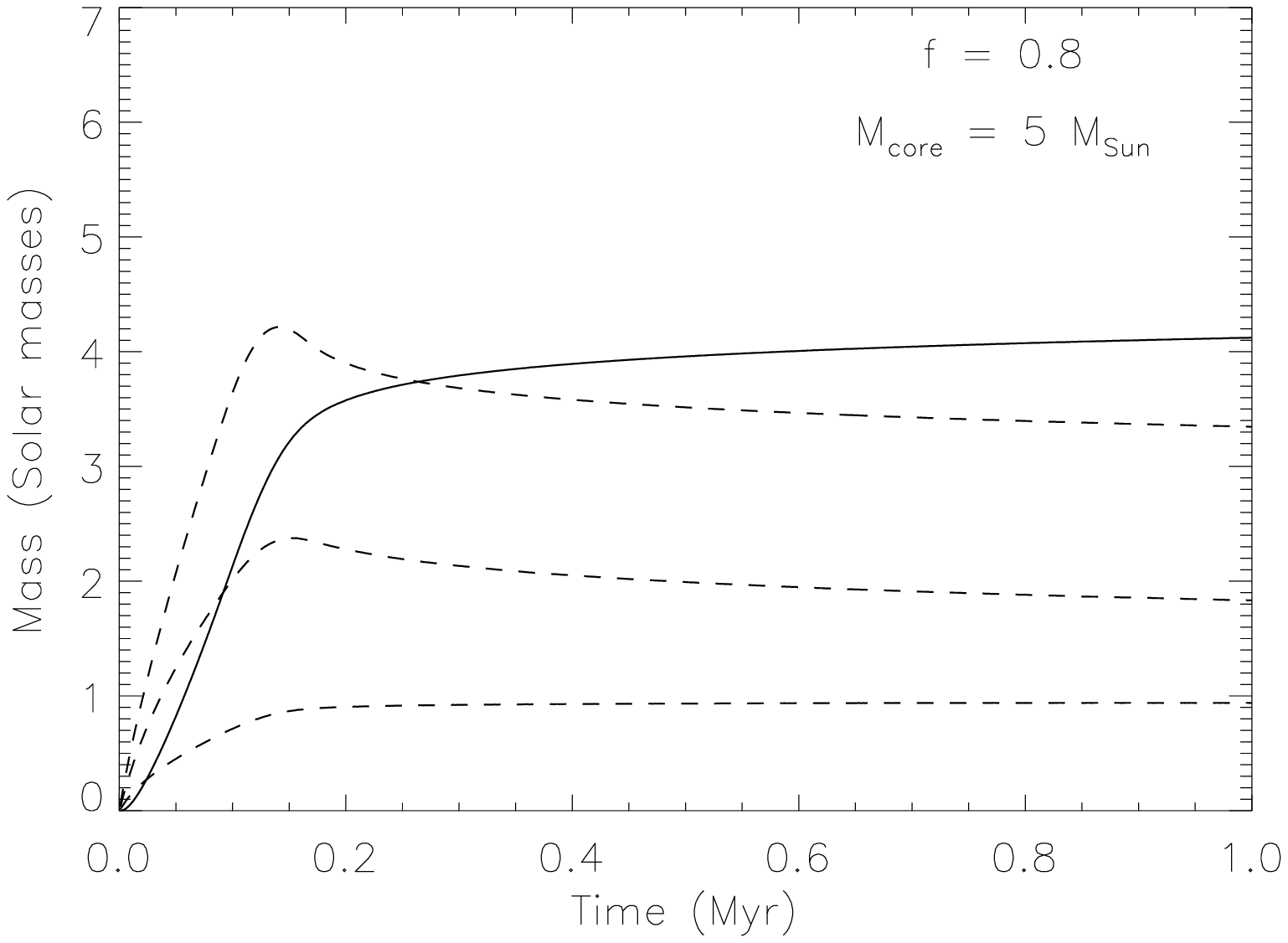,width=0.5\textwidth}
\end{center}
\caption{Protostar (solid line) and disc (dashed lines) masses plotted against time for $f = 0.8$ and for 
$M_{\rm core} = 0.5$ M$_\odot$ (top panel), $M_{\rm core} = 2$ M$_\odot$ (middle panel), and $M_{\rm core}
= 5$ M$_\odot$. The different
dashed lines show the disc mass inside $10$ au, $100$ au, and the total disc mass.  
The figure shows that a significant fraction ($\sim 50$ \%) of the mass inside $100$ au is located in 
the optically thick inner $10$ au. Considering only the disc mass inside $100$ au (since the outer disc 
is potentially susceptible to fragmentation) the disc-to-protostar mass ratio decreases with increasing core mass
which is consistent with observations in the Orion Nebula cluster \citep{eisner08}, and could help to explain
why discs around relatively massive stars disappear on much shorter timescales than discs around lower-mass
TTauri stars.}
\label{Disc_protostar_Mc05Mc2Mc5}
\end{figure}

Figure \ref{Disc_protostar_Mc05Mc2Mc5} is the same as Figure \ref{Disc_protostar_Mc1} except it is for
$f = 0.8$ and $M_{\rm core} = 0.5$ M$_\odot$ (top panel), $M_{\rm core} = 2$ M$_\odot$ (middle panel)
and $M_{\rm core} = 5$ M$_\odot$ (bottom panel).
It is clear from both Figure \ref{Disc_protostar_Mc1} and Figure \ref{Disc_protostar_Mc05Mc2Mc5} that in all cases, 
there is an epoch when the 
disc mass exceeds the mass of the central object. For small $f$ values, this epoch is quite short ($< 1 t_{\rm ff}$)
while for larger $f$ values, the disc may remain more massive than
the central star for many freefall times.  The outer parts ($r > 50 - 100$ au) are, however, susceptible
to fragmentation and may break up to produce companions. The inner parts ($r < 100$ au) of these discs are
only generally comparable in mass to the central
protostar for $t < 2 t_{\rm ff}$.  
This does, however, suggest that our local approximation may not
be strictly valid, and that the effective gravitational $\alpha$ will not be determined by the local
cooling rate. It has been suggested \citep{kratter08, voro09} that there should be two self-gravitating $\alpha$
parametrisations, one when the disc-to-star mass ratio is small, and the other when the disc-to-star mass ratio
is large. This may be the case, but we assume that even if the disc is massive relative to the central star, the
system would still settle into a state with constant $Q$
and be in thermal equilibrium. We don't know if, in such a state, the local dissipation rate will be set by
the local cooling rate, but we assume that such an assumption is at least reasonable at this stage.
The figures also illustrate that in a quasi-steady, self-gravitating state at least 50 \% of the
disc mass within 100 au is located inside 10 au. This large amount of mass in the inner disc could have implications
for, and could aid, planet formation. Detecting such massive inner discs is, however, extremely difficult as they
are very optically thick.  

As discussed in the previous section, it is also likely that 
for $f \ge 0.25$ the outer disc will be very unstable and may fragment to produce companions. The exact radius
at which this occurs depends on the mass of the central star, but is typically between $60$ and $100$ au. 
A significant fraction of the material in the outer disc ($r > 100$ au) is therefore likely to be converted
into companions which will truncate the outer disc at a radius of $50 - 75$ au, depending on where the
fragmentation occurs.  A standard way to determine disc masses observationally is to fit the mass by
modelling the spectral energy
distribution (SED) \citep{robitaille06, andrews09}. The results from such modelling can,
however, be ambiguous \citep{eisner05} and, in particular, a lot of mass can be hidden in optically
thick regions of the disc. The enhanced mass in the optically thick inner $10$ au should therefore
not contribute significantly to the SED, 
and the discs in Figures \ref{Disc_protostar_Mc1} and \ref{Disc_protostar_Mc05Mc2Mc5} would consequently appear to have a mass - determined
through SED modelling - roughly equal to the difference between the mass within $100$ au and the
mass inside $10$ au. For TTauri-like protostar masses (e.g.,  Figure \ref{Disc_protostar_Mc1}),
this would be $0.1 - 0.2$ M$_\odot$ consistent with that obtained for class I sources \citep{eisner05}. 
 
If we only consider the disc mass within $100$ au, since the outer disc is likely
to be removed through fragmentation, Figure \ref{Disc_protostar_Mc05Mc2Mc5} and the bottom panel
of Figure \ref{Disc_protostar_Mc1} also shows that the 
disc-to-protostar mass ratio increases with decreasing
protostar mass, consistent with observations in the Orion Nebula cluster that massive discs
are more common around lower-mass stars \citep{eisner08}.
For core masses below $\sim 1$ M$_\odot$, however, the disc mass is very similar to the mass
of the central star and again brings into question our assumption that the transport 
through self-gravity can be treated as a local process. The inner fragmentation radius 
for these lower-mass cores (see Figure \ref{frag_rad}) is,
however, also closer to $50$ au than $100$ au potentially converting more of the mass in 
the disc into companions.  
The relatively low disc-to-star mass ratio for protostars with masses above a few solar masses
may also help to explain why most discs around relatively massive stars disappear on much
shorter timescales than discs around solar-like TTauri stars. 

\subsection{Additional sources of viscosity}
The discussion in the section above suggests that fragmentation in the outer disc and the presence of an optically thick 
inner disc may result in very young protostars with TTauri-like masses
appearing to have disc masses comparable to that observed.  Figure \ref{Mdot_all} and Figure
\ref{Mdot_Mc1}, however, show that at $t = 1$ Myr, the accretion rates around these protostars is between
$10^{-8}$ and $10^{-7}$ M$_\odot$ yr$^{-1}$.  At this rate, removing the final disc material ($M_{\rm
disc} > 0.3$ M$_\odot$ within $100$ au according to Figures \ref{Disc_protostar_Mc1} and \ref{Disc_protostar_Mc05Mc2Mc5}) would take in
excess of $10^{7}$ yrs.  This is significantly longer than the disc dispersal timescale of 
$\sim 5$ Myr suggested by observations of disc fractions in star forming regions of different ages \citep{haisch01}. 
\citet{rice09} have also shown that the mass accretion rate also drops dramatically
with disc mass, becoming very small for disc masses less than a tenth of the protostar mass. If self-gravity
is the only transport mechanism then the timescale for removing the disc material
would be much longer than observed. 
This suggests that there should be an additional transport mechanism.  The most likely candidate is MRI
operating in the upper layers of the disc \citep{gammie96}.  

\section{Discussion and Conclusions}
We consider here the evolution of protostellar discs that form through the collapse of
molecular cloud cores and in which the primary transport mechanism is self-gravity. 
We assume that the discs settle quickly into a marginally gravitationally stable 
state with $Q = 2$ and are then in thermal equilibrium. In this state angular momentum
transport is driven by an effective gravitational viscosity that is calculated using the 
assumption that energy is dissipated at a rate equal to the rate at which
the disc loses energy through radiative cooling. Following the work of \citet{armitage01}
and \citet{zhu09} we also assume that MRI \citep{balbus91} will operate if the disc is hot enough ($T > 1400$ K)
to be partially ionised. This occurs only in the inner disc and is 
episodic - draining material from the inner disc and then turning off until the inner disc has been
replensished by material from the outer disc - and has therefore been proposed as a mechanism for
explaining FU Orionis outbursts \citep{hartmann96}.

Our simulations consider a range of cores masses ($0.25 - 5$ M$_\odot$) and a range of 
rotation rates ($0.05 \le f \le 1.3$ with $f = \Omega_c / \sqrt{G \rho_c}$).
 The primary results are as follows
\begin{itemize}
\item{The discs settle quickly into a quasi-steady state with a reasonably steep surface density profile
and with a lot ($\sim 50$ \%) of the mass within $100$ au located inside $10 - 20$ au. The
disc-to-star mass ratio decreases with increasing protostar mass, consistent with observations
in the Orion Nebula cluster \citep{eisner08}. Although the disc masses
tend to be higher than observed, with so much mass located in the inner optically thick regions of the disc, 
most of this mass would be difficult to detect with current techniques.}
\item{Although the mass accretion rates are initially higher than those observed, these high accretion rates only
persist for $\sim 1$ free-fall time and quickly drop to values similar to 
that observed.  The simulations also suggest that the accretion rate varies less strongly with protostar mass
($\dot{M} \propto M_*$) than suggested by observations ($\dot{M} \propto M_*^2$) 
\citep{muzerolle05, natta06}, but is consistent with 
the upper boundary of the observed accretion rates \citep{hartmann06}.}
\item{In none of our simulations did the inner disc ($r < 40$ au) have conditions suitable for fragmentation 
($\alpha_{\rm g} > 0.06$). In some cases, the outer disc was susceptible to fragmentation, with the primary factor
determining if a disc could fragment being the rotation rate of the molecular core. The large 
amount of mass in these discs, however, suggests that fragmentation is more likely to result
in sub-stellar or stellar companions that planetary mass objects \citep{kratter09}.
None of our simulations with $f \le 0.1$ satisfied the fragmentation conditions while it was
satisifed for all those
with $f \ge 0.25$.
The inner radius of the fragmentation region increased from $\sim 40$ au for
$M_{\rm core} = 0.25$ M$_\odot$ to $\sim 100$ au for $M_{\rm core} = 5$ M$_\odot$ as did 
the radial range over which fragmentation could occur,
although the outer regions of some of these discs could be stabilised against fragmentation
by cosmic ray ionisation \citep{rafikov09}.}
\item{The mass accretion rate depends strongly on the disc mass
and in all cases drops below $10^{-7}$ M$_\odot$ yr$^{-1}$ when the disc
mass is still quite substantial ($M_{\rm disc} > 0.3$ M$_\odot$). If self-gravity were
to remain the primary transport mechanism, disc clearing timescales would be significantly
longer than that observed. This suggests that an additional tranport mechanism, such
as MRI occuring in the upper layers of the disc \citep{gammie96}, must also operate.  We propose that this
additional mechanism is likely to be negligible initially and become more effective with time. 
This is probably the case for layered accretion which should become more efficient as the disc mass 
decreases and is consistent with the requirement that smalls dust grains must be depleted before
MRI can operate effectively \citep{sano00,sano02,ilgner06}, but the 
pile-up of material in the inner regions of the disc (ultimately resulting in episodic 
FU Orionis-like outbursts) would also not
occur if the additional transport mechanism were too efficient at early times.}
\end{itemize}  

Ultimately it appears that transport driven by self-gravity can explain protostar formation
and disc evolution at early times. 
Our simulations
suggest that disc masses may - at early times - be higher than suggested by observations
but these
large disc masses also suggest that an additional transport mechanism must dominate at 
later times ($t > 1$ Myr) to remove the remaining disc material within observed disc lifetimes.
The pile-up of material in the inner regions of the disc - which may explain FU Orionis
outburst - may also play an important role in the subsequent formation of planets.

\section*{acknowledgements}
P.J.A. acknowledges support from the NSF (AST-0807471), from
NASA's Origins of Solar Systems program (NNX09AB90G), and from NASA's
Astrophysics Theory and Fundamental Physics program (NNX07AH08G).
W.K.M.R. acknowledges support from the Scottish Universities Physics Alliance (SUPA).
The authors would also like to thank the Isaac Newton Institute for Mathematical
Sciences for their hospitality during the Dynamics of Discs and Planets Programme,
and would like to acknowledge useful discussions with Lee
Hartmann, Dick Durisen, Giuseppe Lodato and Neal Turner.

\end{document}